\documentclass{article}

\usepackage[english]{babel}

\usepackage[letterpaper,top=2cm,bottom=2cm,left=3cm,right=3cm,marginparwidth=1.75cm]{geometry}

\usepackage{xcolor}

\newif\ifnotesw \noteswtrue

\usepackage{amsmath, amsfonts}
\usepackage{graphicx}
\usepackage[numbers]{natbib}
\usepackage[dvipsnames]{xcolor}
\usepackage[colorlinks=true, allcolors=blue]{hyperref}
\usepackage{booktabs} 
\usepackage{multicol}
\usepackage{authblk}
\usepackage{float}
\usepackage{subfiles}
\usepackage{algorithm}
\usepackage{algpseudocode}

\newcommand{\yin}{\ensuremath{\mathbf{y}^{\text{in}}_t}}
\newcommand{\yout}{\ensuremath{\mathbf{y}^{\text{out}}_t}}

\title{Leveraging Synthetic and Genetic Data to Improve Epidemic Forecasting}
\author[1]{Dave Osthus}
\author[1]{Alexander C. Murph}
\author[2]{Emma E. Goldberg}
\author[1]{Lauren J. Beesley}
\author[2]{William M. Fischer}
\author[3]{Nidhi K. Parikh}
\author[3]{Lauren A. Castro}
\affil[1]{Statistics Group, Los Alamos National Laboratory}
\affil[2]{Theoretical Biology and Biophysics Group, Los Alamos National Laboratory}
\affil[3]{Information Systems and Modeling Group, Los Alamos National Laboratory}

\begin{document}
\maketitle

\begin{abstract}
Forecasting infectious disease outbreaks is hard. 
Forecasting emerging infectious diseases with limited historical data is even harder.
In this paper, we investigate ways to improve emerging infectious disease forecasting under operational constraints. 
Specifically, we explore two options  likely to be available near the start of an emerging disease outbreak: synthetic data and genetic information.
For this investigation, we conducted an experiment where we trained deep learning models on different combinations of real and synthetic data, both with and without genetic information, to explore how these models compare when forecasting COVID-19 cases for US states. 
All models are developed with an eye towards forecasting the next pandemic.
We find that models trained with synthetic data have better forecast accuracy than models trained on real data alone, and models that use genetic variants have better forecast accuracy compared to those that do not. 
All models outperformed a baseline persistence model (a feat only accomplished by 7 out of 22 real-time COVID-19 cases forecasting models as reported in \cite{lopez2024challenges}) and multiple models outperformed the COVIDHub-4\_week\_ensemble.
This paper demonstrates the value of these underutilized sources of information and provides a blueprint for forecasting future pandemics. 
\end{abstract}

\section{Introduction}
\label{sec:intro}
Over the past decades, infectious disease forecasting has transformed from an academic curiosity into a critical tool for public health preparedness and response. 
This growth has been driven by advances in modeling \citep{brooks2015flexible, osthus2021multiscale, ray2025flusion} and data availability \cite{dong2020interactive}, and by the recognition that timely forecasts can guide resource allocation, inform policy, reduce the cost burden associated with emerging health threats, and improve public health outcomes \citep{ncsl2023}. 
Yet forecasting remains inherently difficult:
challenges include noisy and incomplete data \cite{shadbolt2022challenges}, lags in reporting \cite{beesley2022addressing}, reflexive human behavior \citep{lyu2023human}, uneven policy decisions \citep{kapitsinis2020underlying}, and pathogen evolution \cite{korber2020tracking,beesley2023sars}. 
These obstacles are particularly pronounced during rapidly evolving outbreaks, where even the best models struggle to keep pace \cite{lopez2024challenges}. 
The COVID-19 pandemic brought these struggles into acute focus \cite{ioannidis2022forecasting}. 
On one hand, it marked an unprecedented mobilization of forecasting talent and infrastructure \cite{cramer2022united, sherratt2023predictive}; on the other, it revealed gaps---particularly in integrating real-time signals, quantifying uncertainty, and anticipating the emergence of new viral dynamics.

Among the success stories of the COVID-19 response was the rapid and widespread adoption of genomic surveillance \cite{ling2022challenges}. 
SARS-CoV-2 genomes were sequenced and shared globally at unprecedented scale and speed, offering a near real-time continuously updated feed of the virus’s evolution \cite{shu2017gisaid,korber2025}. 
With relatively low technical barriers and declining sequencing costs, genomic surveillance is poised to remain a cornerstone in future outbreaks. 
Crucially, it enabled the early identification of new variants, which often preceded observable surges in cases \cite{du2023incorporating}. 
This makes variant tracking a potential leading indicator---a rare and powerful feature in infectious disease forecasting.
Looking ahead, this stream of high-resolution, biologically meaningful data offers a promising direction for improving forecast model accuracy.

In general, forecasting systems perform best when three conditions are met: (1) the system’s underlying mechanisms are well-characterized, (2) there is sufficient historical data to train models, and (3) future trends don't deviate too significantly from past ones \cite{hyndman2018forecasting}.
Emerging infectious diseases typically violate the first two conditions, and sometimes all three. 
While mechanistic models---such as compartmental models \cite{brauer2008compartmental}, agent-based models \cite{tracy2018agent}, or phylodynamic models \cite{featherstone2022epidemiological}---can capture transmission and/or evolutionary dynamics, they are often difficult to fit to incomplete and noisy real-time data, and have often been outperformed by their more flexible statistical or machine learning model counterparts in real-time forecasting exercises \cite{mcgowan2019collaborative}.
To perform well, however, these high-potential machine learning forecasting models require substantial training data---data rarely available at the onset of an outbreak from a novel pathogen. 
In the absence of adequate training data, a machine learning model's flexibility can be its weakness, resulting in nonsensical forecasts. 

In this paper, we seek to develop forecasting models that are accurate, scalable, and easily deployed in a pandemic setting where little to no historical data are available for the pathogen of interest.
We consider transformer-based deep learning models for forecasting because, although often costly to train, they are cheap to deploy, can maintain strong performance on new data without frequent retraining; they allow scaling to many data subsets (e.g., geographies), and offer high performance ceilings.

The conspicuous issue for forecasting emerging infectious diseases using deep learning models? Training data.
To learn, these models require large amounts of training data, yet for an emerging pathogen (by definition), datasets are limited.
In this setting, essentially two types of training data are available: historical outbreak data from other pathogens, and synthetic data. 
Neither data source will perfectly represent the pathogen of concern, yet both are available \textit{at outbreak onset} and so can be used to train deep learning models in real-time. 

Historical outbreak datasets are finite, restricted to outbreaks that have occurred and were measured, and thus represent only a portion of the space of possible outbreaks.
Recent work has demonstrated success in incorporating data from different pathogens and surveillance streams to improve forecasting \cite{ray2025flusion, roster2022forecasting}; this strategy is enabled by the public dissemination of public health data \cite[e.g.,][]{project_tycho}.
Thus, while historical data do not represent the emerging pathogen (the forecast target), they do represent ostensibly relevant data, including data reporting vagaries, useful for model training.

Synthetic datasets address many of the shortcomings of historical data.
They are infinite in number (in principle): their generation is constrained primarily by compute resources.
Furthermore, they can represent a diversity of outbreaks limited only by the choice of simulation parameters; measurement noise and biases can added separately to mimic realistic surveillance systems.
Researchers have recently shown success of models trained exclusively on synthetic outbreak data \citep{dudley2025mantis, murph2025synthetic, epifforma}.
This being said, the realism of these outbreaks and the potential gains of using synthetic data in modeling are restricted by the fidelity of the simulator and by the measurement error processes. 

Synthetic data must conform to real or prospective data as it is (or will be) measured. 
Many infectious disease models based on first principles (e.g., compartmental models or agent-based models) can straightforwardly generate time series of case-counts, hospitalizations and deaths.
To make use of viral genetic variant information, as we do here, a synthetic infectious disease simulator must model outbreaks at the variant level, and aggregate variant data to produce ``observed case-count'' time series.
In this paper we make use of one such simulator,  \textit{MutAntiGen} \cite{koelle2015} (discussed in detail in Section \ref{subsubsec:mutantigen}).
This simulator produces both time series of total observed cases (referred to as total cases, or TCs) as well as time series of each constituent variant (referred to as variant-attributable cases, or VACs).

Given this framing, we seek in this work to answer the following five research questions:
 \begin{enumerate}
 \item[\textbf{Q1:}] \emph{Does training with real data or synthetic data produce better forecast performance?}
 \item[\textbf{Q2:}] \emph{Does joint training with real and synthetic data improve COVID-19 case forecasts relative to training with either source individually?}
 \item[\textbf{Q3:}] \emph{Two questions comparing synthetic TC and VAC training data:}
 \begin{enumerate}
 \item[\textbf{Q3a:}] \emph{Does training with synthetic VAC data improve COVID-19 forecasts relative to training with synthetic TC data?}
  \item[\textbf{Q3b:}] \emph{Do models with matched training data and input data outperform models with mismatched training data and input data?}
 \end{enumerate}
 \item[\textbf{Q4:}] \emph{Does including SARS-CoV-2 variant information improve COVID-19 case forecasts?}
 \item[\textbf{Q5:}] \emph{How do these forecasts compare to real-time COVID-19 case forecasts?}
 \end{enumerate}
 
To answer these five questions, we fit and compare eight forecasting models that differ in both the data used to train the models and the inputs to the models, described in Table \ref{tab:models}.
While eight models are defined in Table \ref{tab:models}, they correspond to four different fitted deep learning models trained on different data sets (real, synthetic TCs, synthetic VACs, or real plus synthetic), each applied to two different model input types (TCs or VACs).
For concreteness, models M(r,t) and M(r,v) are using the same fitted deep learning model (the model trained only with real training data) but M(r,t) predicts total cases directly and M(r,v) forecasts each variant-attributable case directly and sums up the individual forecasts (details in Section \ref{subsec:forecasting}).

\begin{table}[H]
\centering
\caption{Forecast model configurations by training data source and input time series. The model naming convention is ``M(training data, input type)" where training data can be r = real, st = synthetic total cases (TCs), sv = synthetic variant-attributable cases (VACs), and a = all sources (real plus synthetic total cases plus synthetic variant-attributable cases) and input types can be t = total cases  and v = variant-attributable cases.}
\label{tab:models}
\vspace{0.5em}  
\begin{tabular}{lll}
\toprule
\textbf{Model} & \textbf{Training Data} & \textbf{Input Time Series} \\
\midrule
\textbf{M(0)} & N/A (persistence baseline) & N/A \\ \midrule
\textbf{M(r,t)} & Real  & TC \\
\textbf{M(st,t)} & Synthetic, TC  & TC \\
\textbf{M(sv,t)} & Synthetic, VAC  & TC \\
\textbf{M(a,t)} & Real + Synthetic & TC \\ \midrule
\textbf{M(r,v)} & Real  & VAC\\
\textbf{M(st,v)} & Synthetic, TC  & VAC \\
\textbf{M(sv,v)} & Synthetic, VAC  & VAC \\
\textbf{M(a,v)} & Real + Synthetic & VAC \\
\bottomrule
\end{tabular}
\end{table}

The models defined in Table \ref{tab:models} allows for a systematic evaluation of how synthetic data and genetic information can improve forecast accuracy by comparing forecast performance of pairs of models. Specifically, 
\begin{itemize}
    \item (\textbf{Q1}) If models trained with synthetic data outperform models trained with real data, we would expect M(st,t) $>$ M(r,t) and M(sv,v) $>$ M(r,v), where M(A) $>$ M(B) means M(A) outperformed M(B).
    \item (\textbf{Q2}) If joint training with real and synthetic data improves COVID-19 case forecasts relative to training with either source individually, we would expect M(a,t) $>$ [M(r,t), M(st,t)], and M(a,v) $>$ [M(r,v), M(sv,v)].
    \item (\textbf{Q3a}) If training with synthetic VACs improves COVID-19 case forecasts relative to training with synthetic TCs, we would expect M(sv,v) $>$ M(st,t). 
    \item (\textbf{Q3b}) If models with matched training data and input data outperform models with mismatched training data and input data, we would expect M(st,t) $>$ M(sv,t) and M(sv,v) $>$ M(st,v).
    \item (\textbf{Q4}) If including SARS-CoV-2 variant information improves COVID-19 case forecasts, we would expect M(r,v) $>$ M(r,t), M(st,v) $>$ M(st,t), M(sv,v) $>$ M(sv,t), and M(a,v) $>$ M(a,t).
\end{itemize}
\noindent Q5 will be evaluated with external models later.

In Section \ref{sec:data}, we describe the scope of the project along with the data used in this exercise. 
In Section \ref{sec:trainingdata}, we present the synthetic data simulator and associated details.
We describe the infectious disease forecasting model in Section \ref{sec:fcst}.
In Section \ref{sec:results}, we present the results of the exercise, including direct answers to all research questions stated above.
Finally, in Section \ref{sec:discussion} we discuss implications, limitations, and future directions of work.

\section{COVID-19 Study Details and Data Overview}
\label{sec:data}
\subsection{Scope of Study}
\label{subsec:scope}
The study details are presented in Table \ref{tab:study}. 
COVID-19 cases were selected as the target because they serve as a leading indicator of more severe outcomes like hospitalizations and deaths and were empirically difficult to forecast \citep{lopez2024challenges}. 
The time range and cadence correspond to data availability and public health relevance.
U.S. states (plus Puerto Rico) were selected as a geographically relevant unit for public health.
There is a large volume of SARS-CoV-2 genomes for US states between June 2020 and December 2022, allowing us to test the value of genetic information.
Furthermore, major COVID-19 data collection and dissemination resources ramped down operation in March of 2023 \cite{jhustopscovid}.

\begin{table}[H]
\centering
\caption{Study details, including the evaluation metrics of mean absolute error (MAE) and weighted interval score (WIS).}
\label{tab:study}
\vspace{0.5em}
\begin{tabular}{@{}ll@{}}
\toprule
\textbf{Category} & \textbf{Details} \\
\midrule
Disease and Target & COVID-19, Cases \\
Time Range & June 2020 – December 2022 \\
Time Cadence & Weekly \\
Geographic Extent & USA \\
Geographic Resolution & States/Territories\\
Forecast Horizon & 1–4 weeks ahead \\
Evaluation Metrics & MAE, WIS, and coverage \\
\bottomrule
\end{tabular}
\end{table}

\subsection{Real COVID-19 Case Data}
\label{subsec:realepi}

\begin{figure}[H]
  \centering
  \includegraphics[width=1\textwidth]{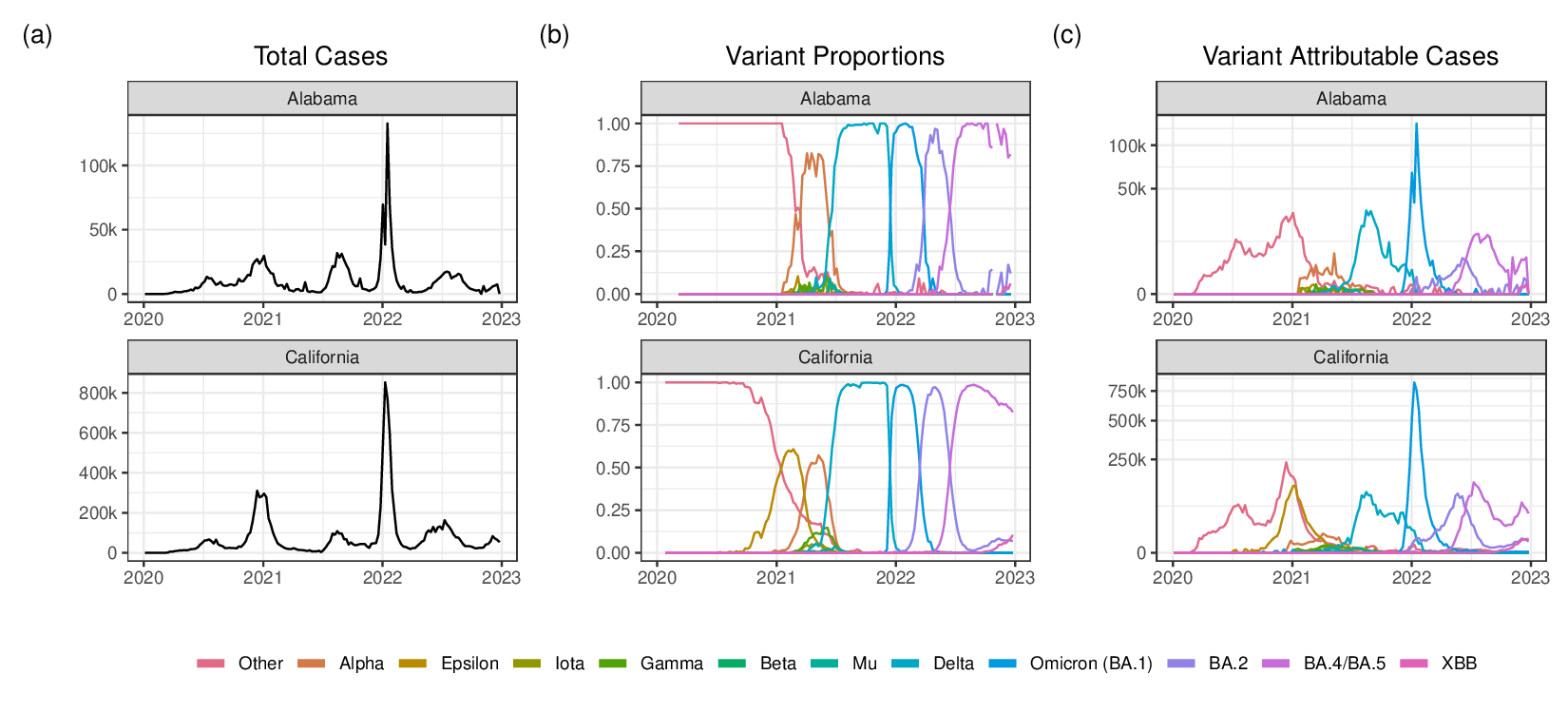}
  \caption{COVID-19 data for Alabama and California. (a) Weekly total cases (TCs). (b) Proportion of sampled viral genomes assigned to each variant. (c) Variant-attributable cases (VACs), computed as TCs times the proportion of genomes assigned to each variant. VACs summed over all variants equal the TCs. Note the square-root scale on the y-axis for better visibility of low-count VACs.}
  \label{fig:ex_covid_data}
\end{figure}

Data on COVID-19 cases between December, 2019 and March, 2023 were obtained from the Johns Hopkins University Center for Systems Science and Engineering (CSSE) GitHub (\url{https://github.com/CSSEGISandData/COVID-19_Unified-Dataset}) \cite{Badu2023}. 
This database includes COVID-19 case data compiled from a variety of sources; our analysis used the case counts reported by the source viewed as the most trustworthy for each location and date. 
Delays in real-time reporting of COVID-19 cases were ignored in this analysis, and the data reported as of September 2023 (the date the data were pulled) for a given date were treated as known as of that date. 
As examples, we show the weekly total COVID-19 case counts for Alabama and California over the study period (Figure \ref{fig:ex_covid_data}). 

\subsection{Real COVID-19 Genetic Data}
\label{subsec:realseq}

The COVID-19 pandemic generated an unprecedented global collection of viral genome sequences, largely coordinated through repositories like GISAID \cite{shu2017gisaid}, which became a central hub for sharing SARS-CoV-2 genomes and metadata. 
In this paper, we made use of approximately 4.5 million sequences gathered in the United States between June 1st, 2020 and December 31st, 2022 available through GISAID.
Rather than using raw genomic sequences, in this work we group sequence variants by Pango lineage designation \cite{o2021assignment} as assigned in the GISAID metadata as of September 2023; each genome is assigned to one of many discrete variant categories, which we aggregate into coherent encompassing supergroups.
Aggregating these labels across time and location yields variant proportion time series, such as those shown in Figure \ref{fig:ex_covid_data}(b).
We combine variant proportion time series with total cases time series to compute variant attributable case (VAC) time series (e.g., Figure \ref{fig:ex_covid_data}(c)), where variant-attributable cases for each time point equal total cases times variant proportion.

We note that there is a meaningful delay between when a viral sample is collected and when its genome and metadata become publicly available.
That lag --- driven by lab turnaround, quality control, metadata completion, and curation --- varies across geography and time \cite{brito2022global}. 
This paper, like many retrospective studies, neglects this delay, but operational forecasting would need to model it.
As such, the results in this paper for the models that use VACs as their input time series should be viewed in their appropriate context.

\section{Training Data}
\label{sec:trainingdata}

\subsection{Real, non-COVID-19 Respiratory Disease Data}
\label{subsec:realtrain}
As early as late 2019, the outbreak later attributed to SARS-CoV-2 was described clinically as an acute respiratory illness (pneumonia) \cite{who2020don229}. 
While little to no COVID-19 data would have been available on January 1st, 2020, we could have known that COVID-19 produced symptoms consistent with respiratory diseases.
For this reason, we use non-COVID-19, real respiratory data available prior to January 1st, 2020 for training in this paper.
All data and code used in this paper are available at \url{https://github.com/lanl/precog/tree/main/synthetic_and_genetic_forecasting}, originally derived from data found at \url{https://github.com/lanl/precog/tree/main/infectious_timeseries_repo}.
Non-COVID-19 respiratory diseases include influenza, pneumonia, mumps, RSV, tuberculosis, and diphtheria (among others).
In this paper, we will often use the shorthand ``real data" or ``real training data" to mean ``non-COVID-19, real respiratory disease data."
For example, the ``Training Data = Real" in Table \ref{tab:models} means ``non-COVID-19, real respiratory disease data."

\begin{table}[H]
\centering
\caption{Training time series summaries. ``Real" means ``non-COVID-19, real respiratory," ``TC" means ``Total Cases," and ``VAC" means ``Variant-Attributable Cases."}
\label{tab:train_ts}
\vspace{0.5em}
\begin{tabular}{@{}cccc@{}}
\toprule
\textbf{Training Data} & \textbf{\# of Time Series} & \textbf{\# of Total Obs.} & \textbf{Avg. (Median) Time Series Length}\\
\midrule
Real & 2,167 & 2,169,760 & 1001 (551) \\
Synthetic, TC & 36,600 & 9,460,880 & 258 (256)\\
Synthetic, VAC & 36,570 & 10,664,618 & 292 (294)\\
\bottomrule
\end{tabular}
\end{table}

\begin{figure}[H]
\centering
\includegraphics[width=1\textwidth]{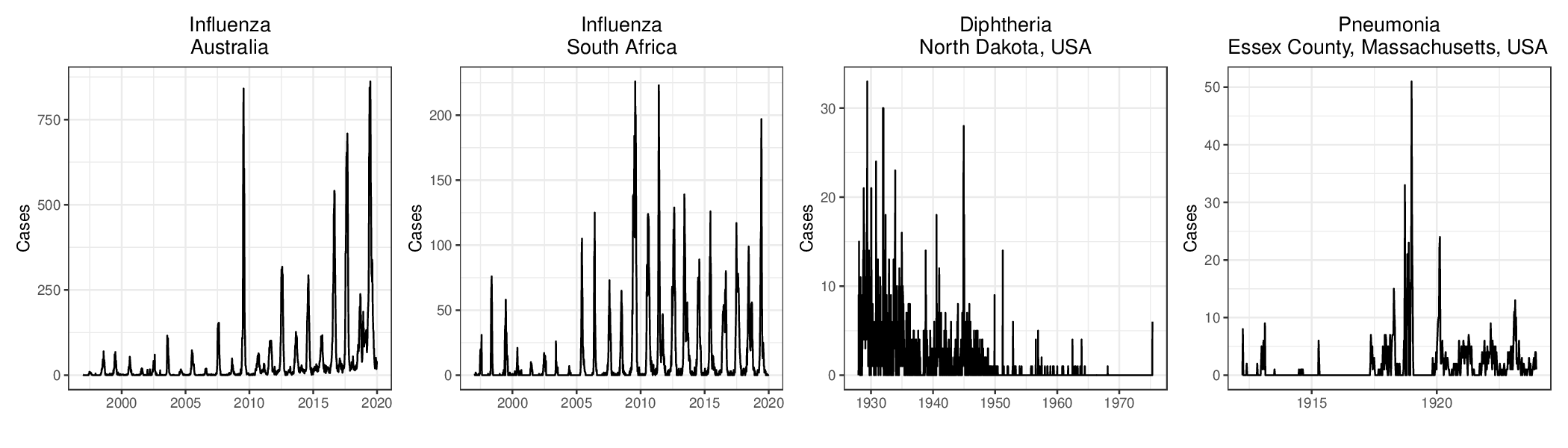}
\caption{Examples of non-COVID-19, real respiratory data. Over 2,000 time series are available for training, amounting to over 2 million observations.}
  \label{fig:ex_noncovid_data}
\end{figure}

As can be seen in Table \ref{tab:train_ts}, about 2,000 non-COVID-19, real respiratory time series are available for training.
Those time series have an average length of about 1000 observations, and a median length of about 550 observations.
A selection of the non-COVID-19, real respiratory time series are shown in Figure \ref{fig:ex_noncovid_data}.

\subsection{Synthetic Data}
\label{subsec:synthetic}

In addition to the real data, we generated synthetic data to represent a wide range of \textit{possible} disease behaviors.  
The intent of these synthetic data is to discover behaviors that are within scope of possible disease dynamics, yet not explicitly represented in the available real data observations.  
We generate synthetic disease data via the MutAntiGen agent-based model (ABM) \citep{bedford2012, koelle2015, castro2020} because it can generate viral variant turnover dynamics.

\subsubsection{MutAntiGen}
\label{subsubsec:mutantigen}

In agent-based modeling, a large-scale ecological system is simulated as a collection of autonomous decision-making entities called agents \citep{bonabeau2002}. In the MutAntiGen ABM, originally developed \cite{koelle2015} as an extension of an antecedent ABM \cite{bedford2012}, ``agents" are categorized as either infected or non-infected individuals in a population susceptible to disease spread. MutAntiGen explicitly models the joint behavior of an evolving pathogen and dynamic susceptibility/resistance of the host population, allowing for non-seasonal case waves. Further details on MutAntiGen are available in the Supplementary Materials Section \ref{S-sec:mutantigendetails}.

\begin{figure}[H]
  \centering
  \includegraphics[width=1\textwidth]{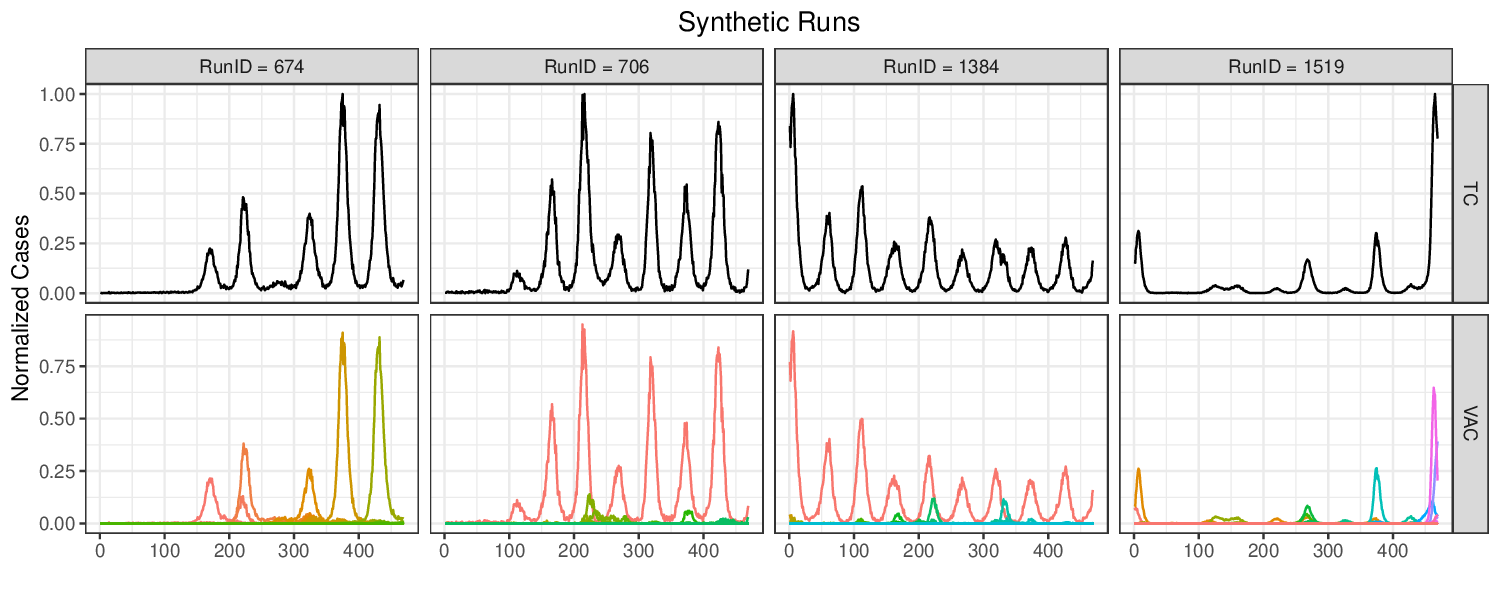}
  \caption{MutAntiGen example runs. MutAntiGen outputs both the total number of cases (top row, TC) and the time series of cases attributed to each variant (bottom row, VAC; each line and color represents a different variant). For each time point, the sum of all variant-attributable cases (bottom row) equals the total cases (top row).}
  \label{fig:ex_synthetic_data}
\end{figure}

Some example runs of MutAntiGen are shown in Figure \ref{fig:ex_synthetic_data}.
In addition to cases over time, the simulator reports viral samples from a subset of infections.
This yields time series of cases attributable to antigenic types, as shown in the lower panels of Figure~\ref{fig:ex_synthetic_data}.  
Viral sampling typically is proportional to number of cases, but this yields very few samples when cases are low before a new wave begins, which is exactly when data are critical for a forecasting model.
We therefore modified the MutAntiGen code to sample with greater intensity when cases are low.

As one might imagine, an ABM intended to represent a massively complicated environmental system includes many parameters to be set by the user prior to running a simulation.  These parameters greatly affect the outcomes of the simulation, but it is still difficult to predict the results of the simulation due to the innate stochasticity of ABMs \cite{gilbert2019}.  The required MutAntiGen parameters and their biological interpretations, as well as our computational modifications to the original code that allow for better sampling at scale, are discussed in Supplementary Materials Sections \ref{S-sec:mutantigendetails} \& \ref{S-sec:mutantigen_coding}.

\subsubsection{Design to Produce MutAntiGen Runs}
Our goal in generating synthetic data was to produce a rich and diverse set of training data for a forecasting model, emphasizing effective representation of both antigenic mutation (an individual property) and turnover in dominant antigenic type (a property of populations). 
Since the aim is to forecast emergent diseases, we are unlikely to know ahead of time what parameter choices will best reflect an impending outbreak; therefore these simulations must cast as wide a net as is practicable, including a broad range of model parameters so that future outbreaks would be more likely to fall within that net (i.e., be represented in the sample space).

To select the parameter values we run MutAntiGen at, we draw a Latin hypercube sample (LHS) \cite{mckay1979}.
Our default MutAntiGen parameter values and ranges were informed by the literature and calibrated to represent global influenza H3N2 phylodynamics. We expanded a subset of the parameters that control the evolutionary, epidemiological, and immunological dynamics to yield more general simulations. For parameters with established empirical estimates, we set plausible bounds based on data from representative RNA viruses including influenza A, influenza B, Measles, Nipah, Dengue, Zika, and Hepatitis C viruses. The specifics of these bounds, our rationales for selecting them, and a full list of MutAntiGen parameters held constant are available in Supplementary Materials Section \ref{S-sec:mutantigendetails}. 

By nature of LHS's independent sampling, many combinations of unrealistic or unknown viruses could be generated, with parameter combinations that do not respect known biological constraints (e.g., error catastrophe, mutation scaling rates \cite{drake1999, holmes2009}) or outbreak conditions ($R_0 > 1$). However, we allowed such combinations under the assumption that biologically nonviable regimes would result in simulation failure or additional data that is harmless.

\subsubsection{Observation Model for Synthetic Data}
\label{subsubsec:modifications}
\begin{figure}[H]
  \centering
  \includegraphics[width=1\textwidth]{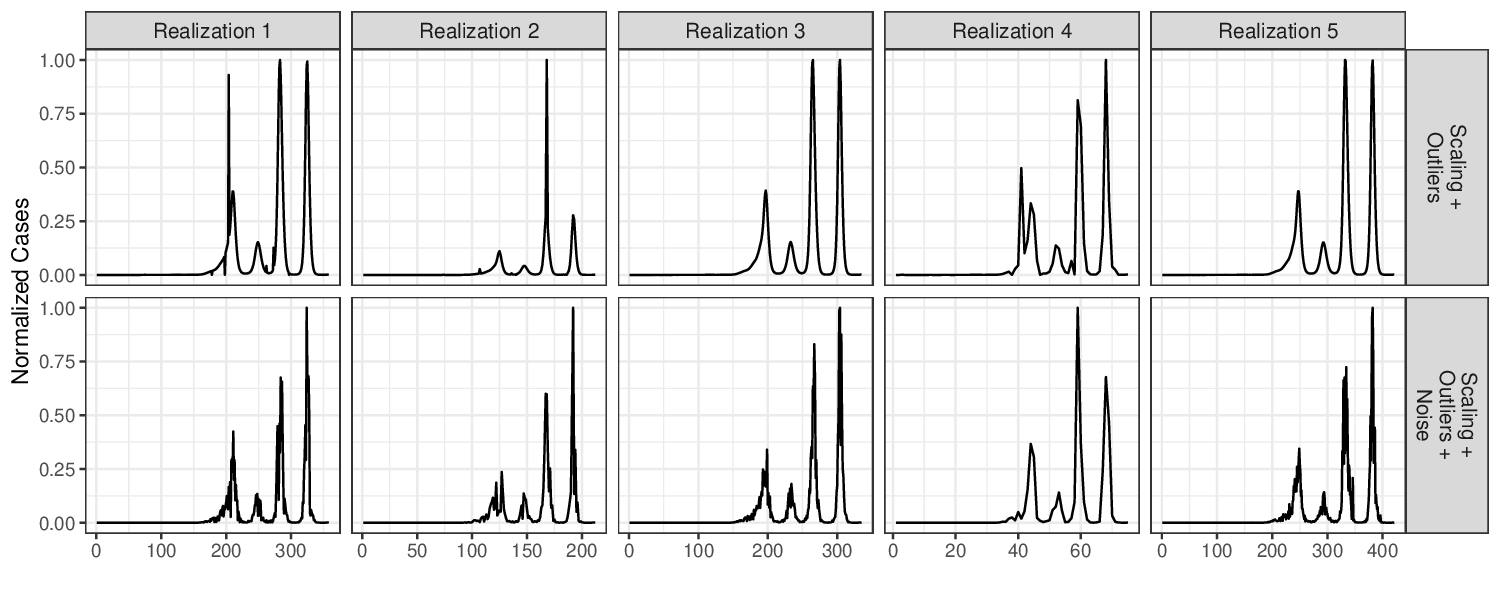}
  \caption{10 of the 20 realizations from the observation model corresponding to a single MutAntiGen output. Realizations were generated by subjecting the ``clean'' MutAntiGen output to either  scaling (random-magnitude compression of the x-axis) and (possible)  addition of outliers (top row), or to scaling plus addition of noise and (possibly) outliers (bottom row).}
  \label{fig:ex_synthetic_data_imperfect}
\end{figure}

As can be seen in Figure \ref{fig:ex_synthetic_data}, the output of MutAntiGen can have unrealistically low noise.
Real data, in contrast, are noisy, biased, and often include outliers (compare Figure \ref{fig:ex_noncovid_data} to Figure \ref{fig:ex_synthetic_data}).
That is, real data can be thought of as imperfect versions of idealized epidemiological data.
In an effort to make the synthetic outputs of MutAntiGen more realistic, we passed each MutAntiGen output through an observation model 20 times, resulting in different imperfect versions of each MutAntiGen time series.
This observation model increases both the amount and the diversity of the training data.
Figure \ref{fig:ex_synthetic_data_imperfect} shows different realizations of the observation model for a single MutAntiGen output.
After sending all MutAntiGen runs through the observation model process, we generated approximately 36,000 total time series for training for both the total cases and the variant-attributable cases (see Table \ref{tab:train_ts} for specifics).
More details of the observation model can be found in Section \ref{S-sec:obsmodel} of the Supplementary Materials.

\section{Forecasting Model}
\label{sec:fcst}

\subsection{Training}
\label{subsec:training}
We frame our probabilistic forecasting problem as a conditional quantile regression problem.
Let
\begin{align}
    y_{t+h}|\yin&\sim F_{\yin,h}
\end{align}
\noindent be the conditional distribution for $y_{t+h}$, the number of new infections reported at time $t+h$ for $t \in \{1,2,\ldots\}$ and $h \in \{1, 2, \ldots, H\}$ given the last $C$ newly reported infections $\yin = y_{(t-C+1):t}$.
In this paper, we set $C = 20$ and $H = 4$.
That is, our 1-, 2-, 3-, or 4-step-ahead forecasts are based on the last 20 observations of the time series when making a forecast at time $t$.

We define the quantile function as follows:
\begin{align}
    q_{\tau,h}(\yin) := F^{-1}_{\yin,h}(\tau)
\end{align}
\noindent for $\tau \in [0,1]$. 
That is, the conditional quantile function $q_{\tau,h}()$ --- doubly indexed by the quantile level $\tau$ and the step ahead $h$ --- is the inverse of the conditional cumulative distribution function evaluated at the quantile level $\tau$, $F^{-1}_{\yin,h}(\tau)$.
We approximate the inverse of the conditional cumulative distribution function $F^{-1}_{\yin,h}$ with a set of quantile functions $q_{\tau,h}(\yin)$ evaluated over a grid of quantile levels $\tau \in \mathcal T$ where 
\begin{align*}
    \mathcal T = \{0.0005, 0.005, 0.01, 0.025, 0.05, 0.1, \ldots, 0.9, 0.95, 0.975, 0.99, 0.995, 0.9995\}
\end{align*} 
\noindent and $|\mathcal T| = 27$.
$\mathcal T$ constitutes a dense grid of quantile levels, denser than we use for evaluation (see Section \ref{subsec:metrics} for details).
This dense grid, however, allows us to better approximate the tails of the forecast distributions which will be needed in the forecasting of models M(r,v), M(st,v), M(sv,v), and M(a,v) (see Section \ref{subsec:forecasting} for details).
The quantile function provides a point forecast and forecast intervals. 
For instance, the point forecast is the evaluated quantile function when $\tau = 0.5$ (the median). 
The 95\% forecast interval lower and upper bounds are the quantile functions evaluated for quantile levels $\tau = 0.025$ and $\tau = 0.975$, respectively.

Given our problem statement, our next task is to estimate $q_{\tau,h}(\yin)$.
To do that, we turn to deep learning \cite{lecun2015deep}.
Deep learning models are flexible function approximators.
Provided an adequate amount of training data, deep learning models can learn continuous functions to high degrees of precision \cite{hornik1989multilayer}.
Using the training data described in Section \ref{sec:data}, we train a 2-layer transformer model \cite{vaswani2017attention} that takes the last 20 observations of a time series as input and predicts the quantile levels in $\mathcal T$ for $h \in \{1, 2, \ldots, H\}$.
Pinball loss is used to perform the quantile regression.
Training and deep learning model details can be found in Section \ref{S-sec:model_details} of the Supplemental Materials.

The results of the model fitting are four different trained deep learning models, differing only in the training data used to learn the model parameters: real data only (i.e., all non-COVID-19, real respiratory data detailed in Section \ref{subsec:realtrain}), synthetic total cases, synthetic variant-attributable cases, and all training data (i.e., non-COVID-19, real respiratory data, synthetic total cases data, and synthetic variant-attributable cases). 
As is detailed in Table \ref{tab:training_summary}, with $C = 20$, we are able to generate between 2 and 21 million input/output pairs of training data where $\yin$ is the input and $\yout = y_{(t+1):(t+H)}$ is the output.
To be clear, if there is a training time series of length $T=100$, that will produce $100 - 20 - 4 + 1 = 77$ input/output training data examples (e.g., $\{\mathbf{y}^{\text{in}}_C,\mathbf{y}^{\text{out}}_C\}$, 
$\{\mathbf{y}^{\text{in}}_{C+1},\mathbf{y}^{\text{out}}_{C+1}\}$, 
$\ldots$, 
$\{\mathbf{y}^{\text{in}}_{T-C-H+1},\mathbf{y}^{\text{out}}_{T-C-H+1}\}$).

\begin{table}[H]
\centering
\caption{Training details for each training data set. Each model was presented with approximately 25 million training examples during learning. Example views is the average number of times each training example was viewed by the model during training (total number of training examples viewed by the model divided by the total number of unique available training examples). The closer example views is to 1 (or less), the less likely the model is to memorize the training data and thus the less likely the model is to overfit the training data. Training time does not scale with the number of training examples but rather the number of training examples presented to the model during training, which was held fixed at approximately 25 million for all models. Training times presented here used a 2023 MacBook Pro with Apple M2 Max, using CPU-only training on 1 CPU core, with 64 GB ram. Code was run in \texttt{R} 4.4.3 with models fit using the \texttt{torch} package (version 0.16.3).}
\label{tab:training_summary}
\vspace{0.5em}
\begin{tabular}{@{}cccc@{}}
\toprule
\textbf{Training} & \textbf{Number of Training} & \textbf{Example} & \textbf{Training Time}\\
\textbf{Data} & \textbf{Examples (millions)} & \textbf{Views}& \textbf{(minutes)}\\
\midrule
Real & 2.1 & 11.8 & 706\\
Synthetic, TC & 8.6 & 2.9 & 701\\
Synthetic, VAC & 9.8 & 2.5 & 701\\
All & 20.6 & 1.2 & 702\\
\bottomrule
\end{tabular}
\end{table}

Table \ref{tab:training_summary} presents summaries of model training. 
Each model was trained on between 2 and 21 million unique training examples.
Each model was presented with approximately 25 million training examples during training.
Thus, each training example was viewed by the model between 1 and 12 times (example views), depending on the number of available training examples. 
Deep learning models run the risk of memorizing the training data and thus overfitting if they are presented the same training examples repeatedly. 
The large numbers of training examples shown in Table \ref{tab:training_summary} help protect against overfitting.
The training time is a function of the number of training examples presented to the model (25 million), not the number of available training examples. 
This is why the training time does not scale with the number of training examples.
The time to train the model (almost 12 hours) would be considered expensive (possibly prohibitive) if retraining was required every week when new data become available for real-time forecasting.
The forecasting approach considered here falls under the ``expensive to train but cheap to deploy" paradigm.
While it takes on the order of half a day to train these models, they only need to be trained once. 
Forecasting with these trained models is measured on the scale of seconds (or less), making them appealing in operational settings.
It is worth noting that the training time could be shortened by making use of graphical processing units (GPUs).

\subsection{Forecasting}
\label{subsec:forecasting}

Forecasting differs depending on the model input (recall Table \ref{tab:models}).
The models that take total cases as the input are straightforward to forecast. 
The fitted transformers produce forecasts for all quantile levels in $\mathcal T$.
We only use seven of those quantile levels, $\tau \in \{0.025, 0.1, 0.25, 0.5, 0.75, 0.9, 0.975\}$, allowing us to produce a point forecast (the median) and three prediction intervals: 50\%, 80\% and 95\%.
These quantile predictions allow us to evaluate forecasts with respect to multiple popular metrics (see Section \ref{subsec:metrics} for details).

Forecasting the models where the model input are the variant-attributable cases (VACs) requires one more step.
For a given state, forecast date ($t$), and forecast horizon ($h$), we forecast each VAC at the dense grid of 27 quantile levels in $\mathcal T$.
Using those 27 quantiles as an estimate of the cumulative distribution function (CDF) for each VAC, we draw a realization from each VAC's CDFs by inverting the CDF \cite{devroye2006nonuniform} and linearly interpolating between quantile estimates (this is why $\mathcal T$ has quantile estimates so far out in the tails).
Given a draw from each VAC's forecast distribution, we sum those draws constituting a single draw from the total cases forecast distribution.
We repeat this sampling process N = 100,000 times.
Then, we compute the same seven quantile levels $\{0.025, 0.1, 0.25, 0.5, 0.75, 0.9, 0.975\}$ as the sample quantiles of the N draws from the total cases forecast distribution, derived from each individual VAC forecast distribution.

Notice that pairs of models in Table \ref{tab:models} use the same fitted transformer model to produce forecasts.
For example, models M(r,t) and M(r,v) each use the same fitted transformer model, trained with only non-COVID-19, real respiratory data, but will yield different forecasts because the inputs to the model are different (recall Figure \ref{fig:ex_covid_data} (a) and (c)).

\section{Forecasting Results}
\label{sec:results}
In Section \ref{subsec:metrics}, we detail the metrics used to perform our evaluation.
In Section \ref{subsec:results}, we provide high-level findings from our exercise.
In Section \ref{subsec:researchqs}, we directly answer the research questions stated in Section \ref{sec:intro}.
\subsection{Evaluation Metrics}
\label{subsec:metrics}
Before jumping into the results, we define the metrics we use to evaluate the forecasts: mean absolute error (MAE), weighted interval score (WIS), and empirical coverage.

MAE is defined as 
\begin{align}
    \text{MAE} &= \frac{1}{N}\sum_{i=1}^N |\hat{y}_i - y_i|
\end{align}
\noindent where $N$ is the number of forecasts, $y_i$ is the observation, and $\hat{y}_i$ is the point forecast. 
MAE is greater than or equal to 0 and is negatively oriented (smaller is better).
The point forecast $\hat{y}$ for this work is the median forecast from the quantile regression.
Most of the evaluation results are presented relative to a persistence model --- a model whose forecast $\hat{y}_{t+h} = y_t$ for any $h \geq 1$ (i.e., M(0)).
As such, we also define relative MAE (rMAE) as follows:
\begin{align}
    \text{rMAE}(\text{M}_i) &= \frac{\text{MAE for M$_i$}}{\text{MAE for M(0)}} 
\end{align}
\noindent where M$_i$ is any model $i$ listed in Table \ref{tab:models}.
rMAE is greater than or equal to 0 and negatively oriented.
rMAE(M$_i$) $< 1$ indicates that model M$_i$ has a better MAE than a persistence model.

WIS is a way to evaluate forecasts in an interval format.
Intuitively, WIS penalizes two things: interval widths (the wider the interval width, the larger the penalty) and observations that fall outside the forecast interval (the further the observation falls outside the forecast interval, the larger the penalty). 
As such, WIS encourages forecasts to be sharp and well-calibrated \cite{gneiting2007probabilistic}.
We consider $K = 3$ central $(1- \alpha) \times 100\%$ forecast intervals: 50\%, 80\% and 95\% (corresponding to $\alpha = 0.5, 0.2, \text{ and } 0.05$, respectively).
Following \cite{bracher2021evaluating}, WIS is defined as 
\begin{align}
    \text{WIS} &= \frac{1}{K+0.5} \times \Bigg( w_0 \times |y-q_{0.5}| + \sum_{k=1}^K (w_k \times \text{IS}_{\alpha_k})\Bigg)
\end{align}
\noindent where $w_0 = 0.5$, $w_k = \alpha_k/2$, $q_{0.5} = \hat{y}$ (the median forecast), $y$ is the observations, and $\text{IS}_{\alpha_k}$ is the interval score corresponding to $\alpha_k$, defined as

\begin{align}
    \text{IS}_{\alpha_k} &= (u_{\alpha_k} - l_{\alpha_k}) + \frac{2}{\alpha_k} \Bigg( (l_{\alpha_k} - y) * \textbf{I}(y < l_{\alpha_k}) + (y - u_{\alpha_k}) * \textbf{I}(y > u_{\alpha_k})\Bigg)
\end{align}
\noindent where $l_{\alpha_k}$ and $u_{\alpha_k}$ are the $\alpha_k/2$ and $1-\alpha_k/2$ quantiles and $\textbf{I}()$ is an indicator function equal to 1 if the argument is true and 0 otherwise.
WIS is greater than or equal to zero and is negatively oriented. 
Similar to MAE, relative WIS (rWIS) is defined as follows:
\begin{align}
    \text{rWIS}(\text{M}_i) &= \frac{\text{WIS for M$_i$}}{\text{WIS for M(0)}} 
\end{align}
\noindent where M$_i$ is model $i$.
rWIS(M$_i$) $<$ 1 means model M$_i$ has a better WIS than a persistence model.
The persistence model ($\hat{y}_{t+h} = y_t$) does not intrinsically produce forecast intervals. 
We follow the procedure described in the Methods Section of \cite{cramer2022evaluation} where the forecast intervals of the persistence model are based on the historical changes of the observed COVID-19 cases time series for each state.

Empirical coverage for a $(1-\alpha)\times 100$\% forecast interval is the proportion of forecast intervals that contain the observation $y$.
Empirical coverage is defined as
\begin{align}
    \text{Coverage}(\alpha) &= \frac{1}{N} \sum_{i=1}^N \textbf{I}(l_{i,\alpha} \leq y_i \leq u_{i,\alpha})
\end{align}
\noindent where $l_{i,\alpha}$ and $u_{i,\alpha}$ are the lower and upper quantile estimates corresponding to the $(1-\alpha)\times 100\%$ forecast interval. 
Probabilistically well-calibrated forecasts are those whose empirical coverages are nearly equal to their nominal coverages (i.e., $1-\alpha$) for all $\alpha$ levels.

\subsection{Overview of Results}\label{subsec:results}

\begin{figure}[H]
  \centering
  \includegraphics[width=1\textwidth]{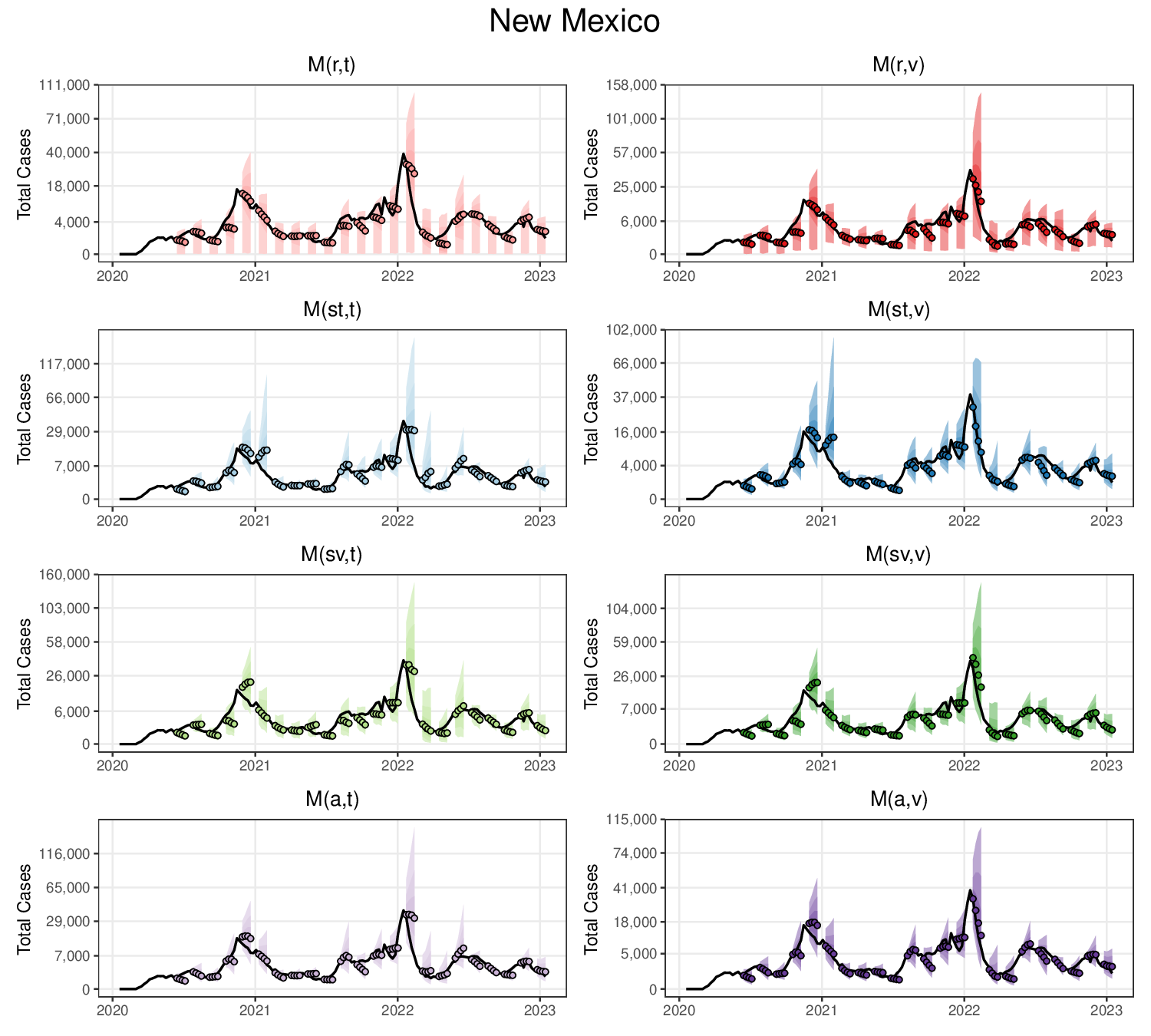}
  \caption{Selected 1-week-ahead through 4-week-ahead forecasts for New Mexico for all models. Black line: total cases time series. Colored points: median forecast cases. Ribbons mark the 50\%, 80\%, and 95\% forecast intervals. Note: y-axis is on a square root scale to better see the low-case-count forecasts.}
  \label{fig:ex_fcst}
\end{figure}

Figure \ref{fig:ex_fcst} shows selected forecasts for New Mexico for all models.
As expected, forecast interval widths increase with increasing horizon $h$.
The forecasts for models trained on only real data (M(r,t) and M(r,v)) have noticeably low quantile estimates at the 0.025 level.
The real data used for training are fairly noisy, which may have led to this forecast behavior.
The other notable trend is the forecasts made at the peak of the Omicron (BA.1) wave in early 2022.
The models that take TCs as input, M(.,t) (left column of Figure \ref{fig:ex_fcst}), produced flat to mildly dropping forecasts at the peak in early 2022, while the models that take VACs as inputs, M(.,v) (right column of Figure \ref{fig:ex_fcst}), correctly produced a steep drop in their forecasts.
Figure \ref{fig:ex_fcst} illustrates that there are systematic differences in the forecast models that break down along training data and model input lines.

\begin{figure}[H]
  \centering
  \includegraphics[width=.49\textwidth]{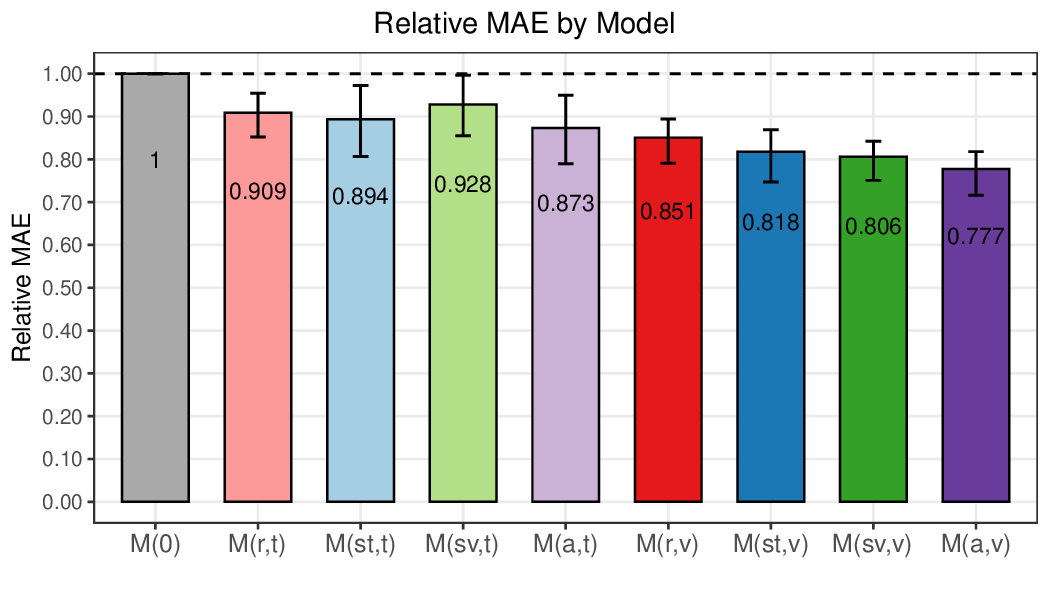}
  \includegraphics[width=.49\textwidth]{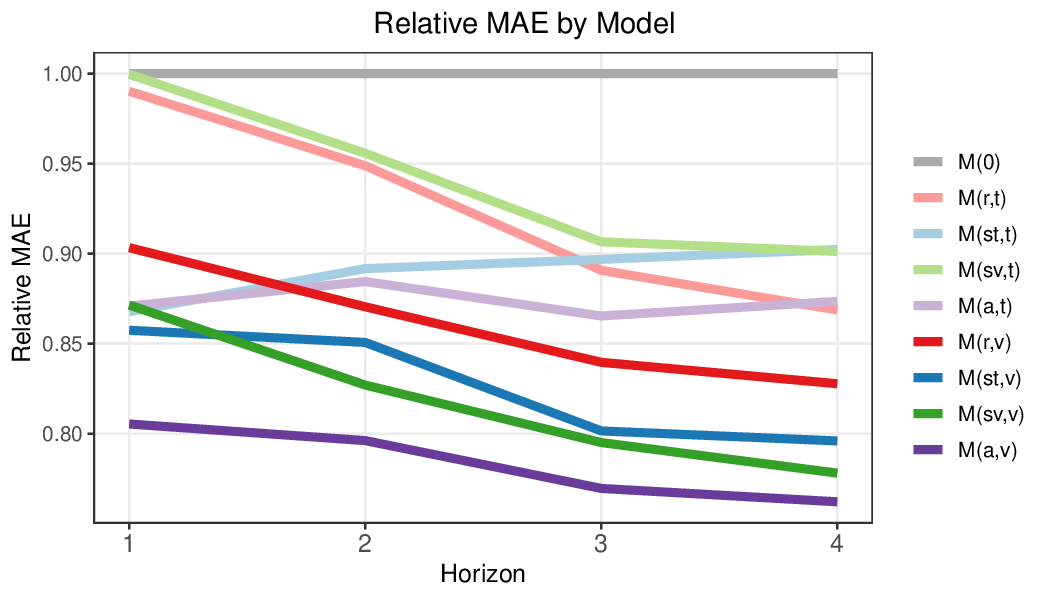}
  \includegraphics[width=.99\textwidth]{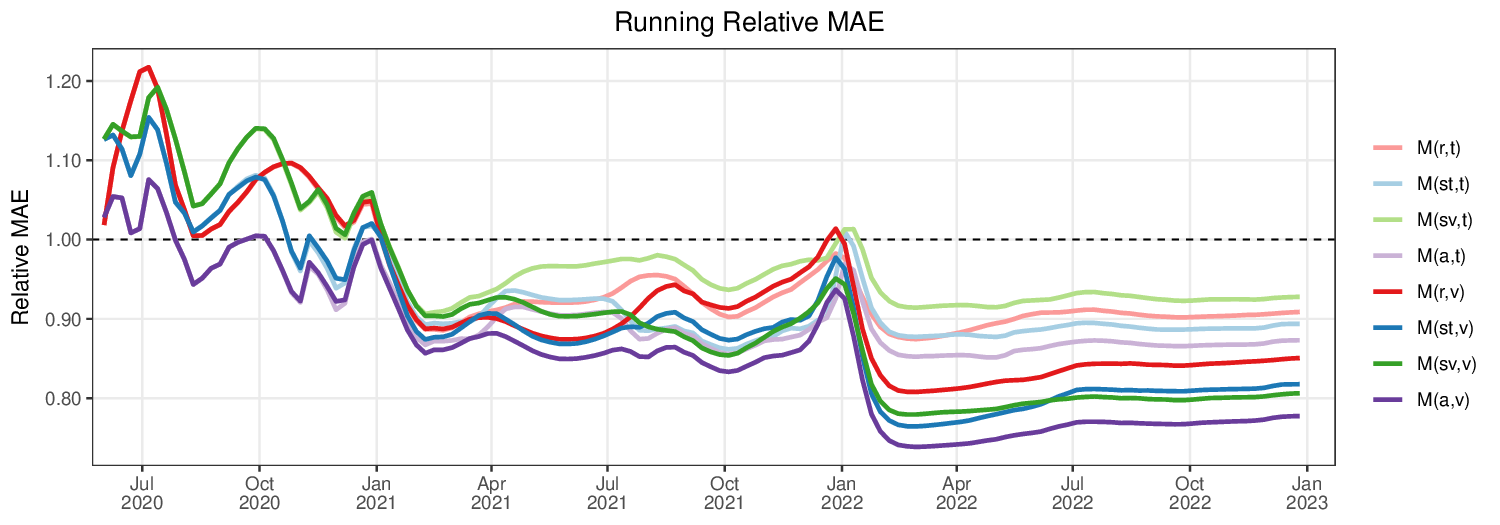}
  \caption{(top left) Overall relative mean absolute error (rMAE) with 95\% confidence intervals as error bars. Error bars were constructed via bootstrapping. (top right) Relative MAE by forecast horizon. (bottom) Running relative MAE. The running relative MAE for each date is the relative MAE if the evaluation period ran from June 1st, 2020 through the x-axis date.}
  \label{fig:mae}
\end{figure}

Figure \ref{fig:mae} shows rMAE results. 
Overall, we see all models had better MAE than the baseline, the worst being M(sv,t) with 0.928 rMAE and the best being M(a,v) with 0.777 rMAE.
Uncertainty intervals are 95\% confidence intervals, obtained via bootstrapping (see Section \ref{S-sec:bootstrapping} for details). 
Each model had rMAE less than or equal to 1 for all forecast horizons.
The bottom of Figure \ref{fig:mae} shows a running relative MAE where the rMAE on a given date corresponds to the evaluation of all forecasts available between June 1st, 2020 through the x-axis date. 
It is clear the results depend on the evaluation period, as models perform differently throughout different phases of the pandemic.
That said, for all evaluation end times after January 2021, all models had an rMAE less than 1 (except for a momentary blip for a few models around January 2022).

\begin{figure}[H]
  \centering
  \includegraphics[width=.49\textwidth]{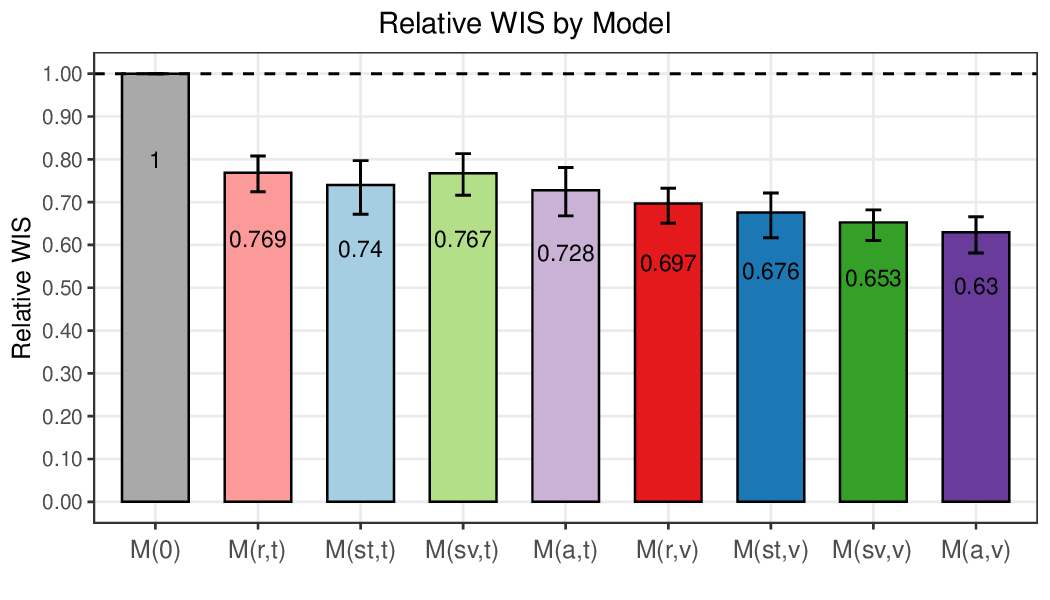}
  \includegraphics[width=.49\textwidth]{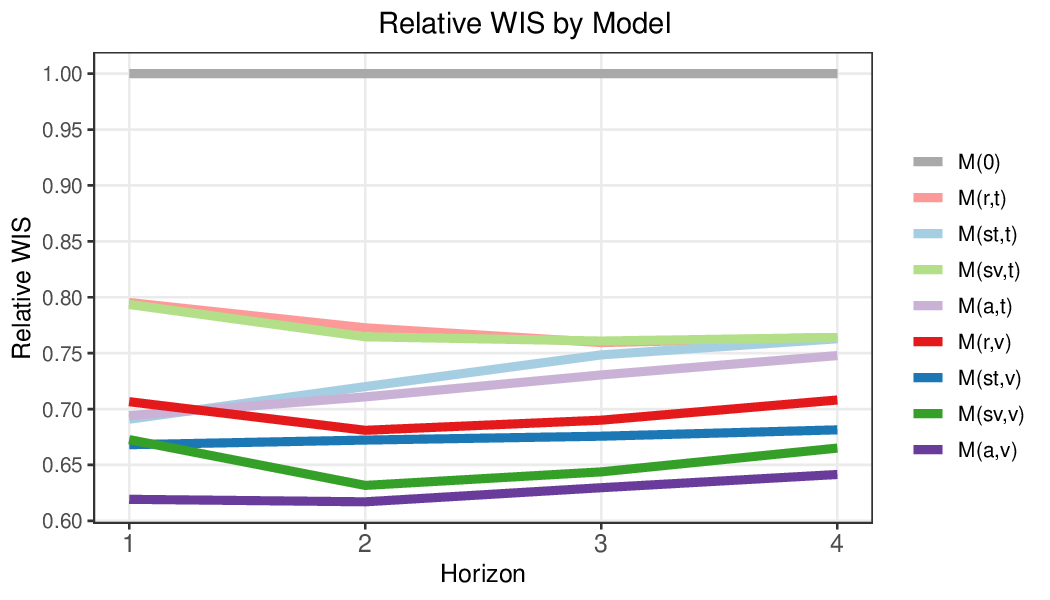}
  \includegraphics[width=.99\textwidth]{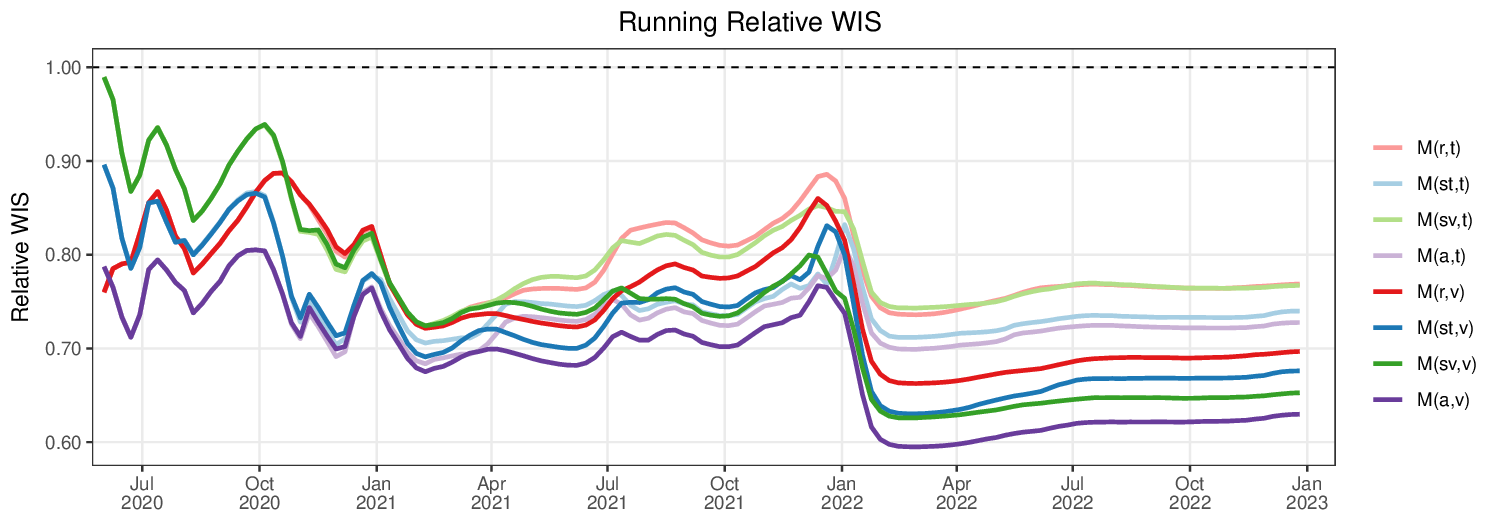}
  \caption{(top left) Overall relative weighted interval score (WIS) with 95\% confidence intervals as error bars. Error bars were constructed via bootstrapping. (top right) Relative WIS by forecast horizon, and (bottom) running relative WIS. The running relative WIS for each date is the relative WIS if the evaluation period ran from June 1st, 2020 through the x-axis date.}
  \label{fig:wis}
\end{figure}

Figure \ref{fig:wis} shows results for WIS relative to a persistence model. 
Overall, all models had rWIS less than 1 ranging from 0.769 (M(r,t)) to 0.63 (M(a,v)).
Each model also had rWIS less than 1 (in fact, less than or equal to 0.80) for all considered forecast horizons.
The running relative WIS is less than 1 for all models and all evaluation end dates between June 2020 through December 2022 providing strong evidence that all models demonstrated better forecast performance as measured by WIS compared to M(0).

\begin{figure}[H]
  \centering
  \includegraphics[width=1\textwidth]{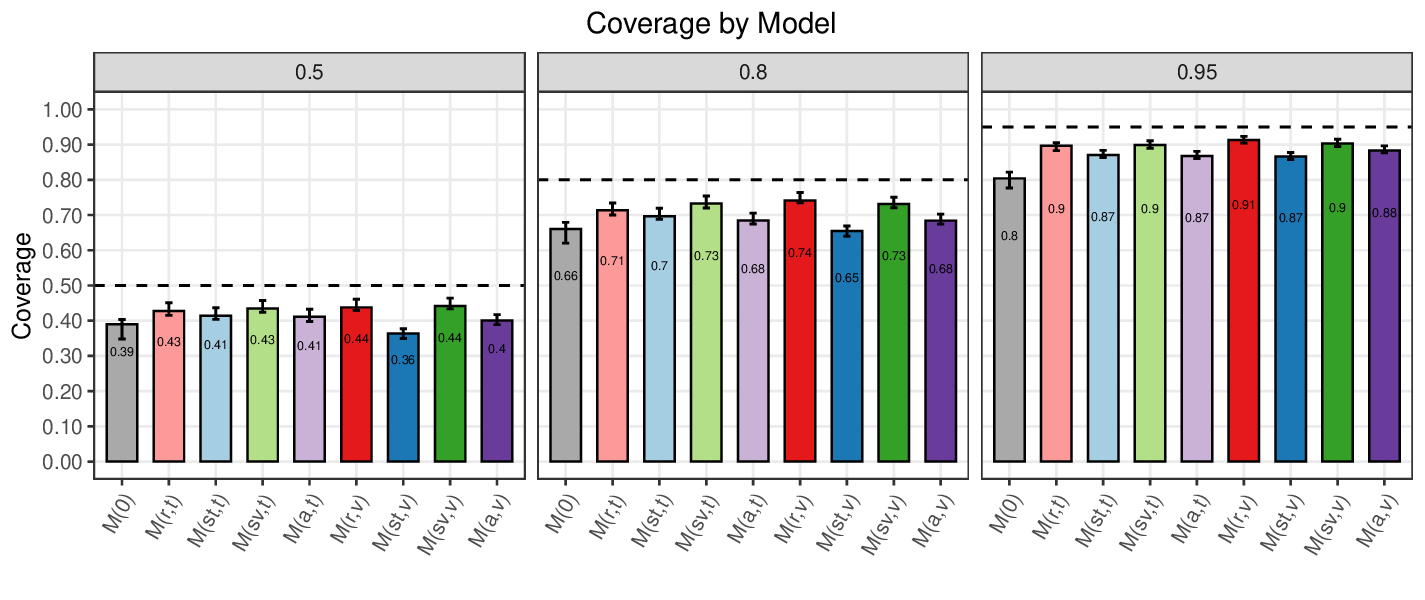}
  \includegraphics[width=1\textwidth]{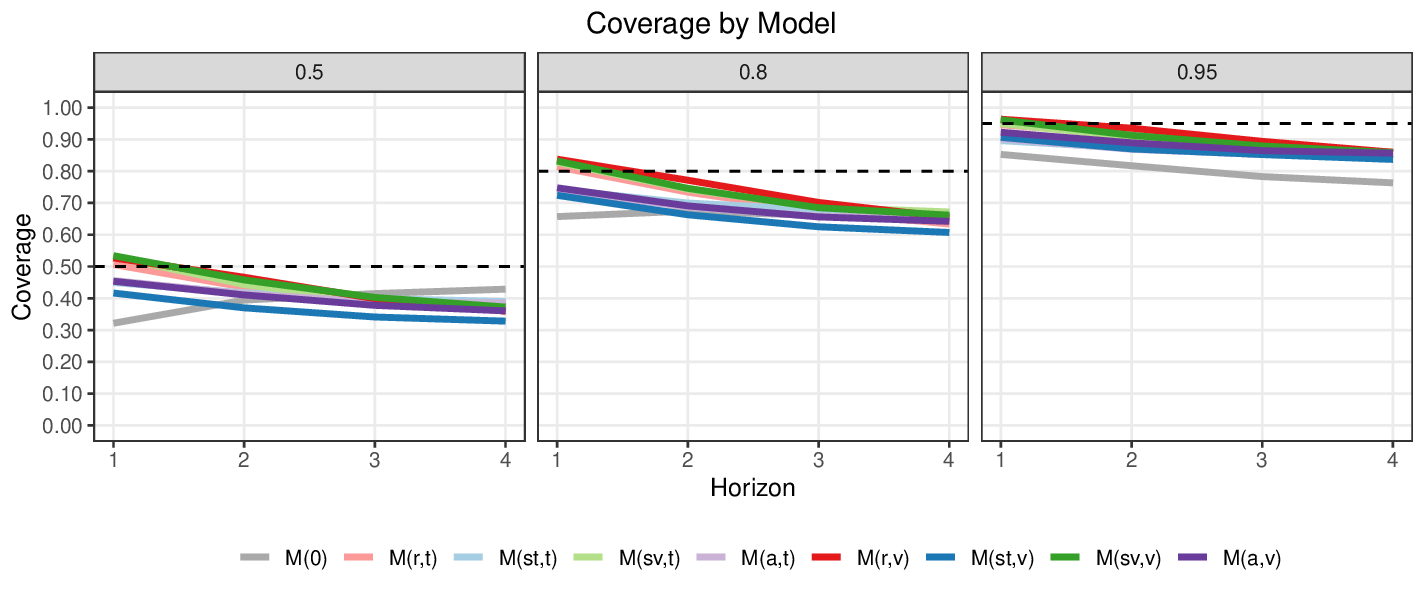}
  \caption{(top) Overall empirical coverages for 50\%, 80\%, and 95\% forecast intervals. (bottom) Empirical coverage by horizon. The dashed horizontal line indicates the nominal coverage. Undercoverage exists for all models and horizons.}
  \label{fig:coverage}
\end{figure}

Figure \ref{fig:coverage} shows the empirical coverage for all models, overall and by horizon. 
All models exhibit undercoverage (i.e., observations fell outside their predictive intervals more often than expected), a common occurrence in COVID-19 forecasting \citep{lopez2024challenges}.
All models showed empirical coverage similar to that of a persistence model for 50\% forecast intervals, but empirical coverages closer to nominal than the persistence model for 95\% forecast intervals.
When we examine empirical coverage by forecast horizon (bottom of Figure \ref{fig:coverage}), we generally see that all non-persistence models have near nominal coverage at horizon 1, but that empirical coverage drifts below nominal coverage as horizon increases to 4 weeks ahead.

\subsection{Answering the Research Questions}
\label{subsec:researchqs}
We now directly answer our research questions laid out in Section \ref{sec:intro}.

\subsubsection{Q1: Does training with real data or synthetic data produce better forecast performance?}
\begin{figure}[H]
  \centering
  \includegraphics[width=1\textwidth]{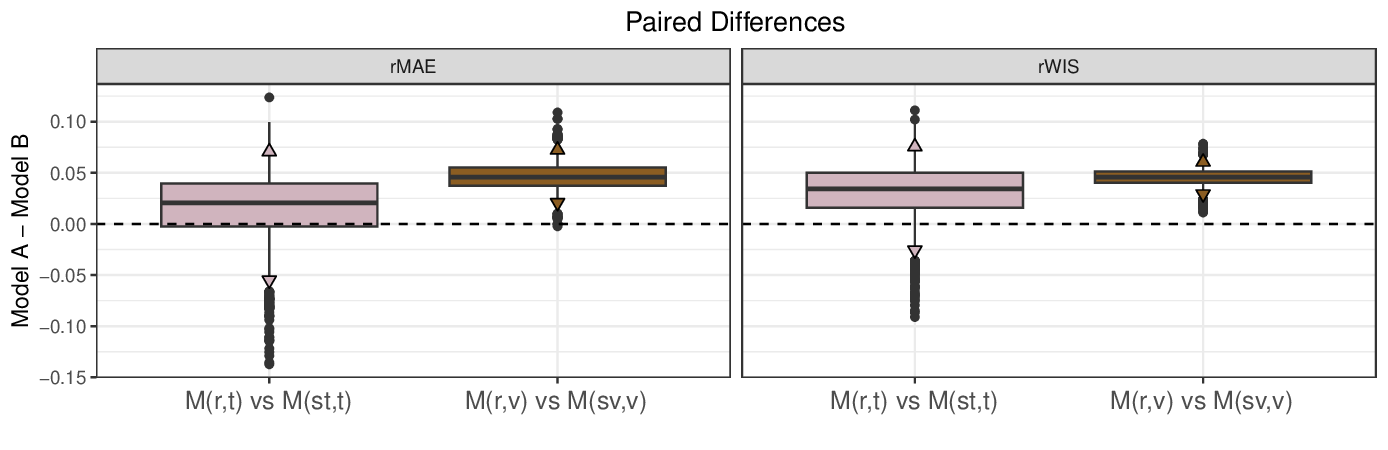}
  \caption{Distribution of 5,000 bootstrapped model differences for rMAE and rWIS. Triangles indicate the 2.5 and 97.5 percentiles. Presented results show Model A - Model B (on the x-axis it is displayed as M(A) vs M(B)). For instance, M(r,t) vs M(st,t) in the rMAE panel presents the results of M(r,t)'s rMAE - M(st,t)'s rMAE, across all bootstrap samples. Positive numbers indicate Model B performed better than Model A. Fairly strong evidence is shown that models trained with synthetic data alone performed better than models trained with real data alone.}
  \label{fig:q1}
\end{figure}

There is fairly strong evidence that training with synthetic data alone produces better forecast performance than training with real data alone. 
That evidence is presented in Figure \ref{fig:q1}, which shows the paired differences of rMAE and rWIS across 5,000 bootstrapped samples. 
The model comparisons were M(r,t) vs M(st,t) and M(r,v) vs M(sv,v).
We see that in over 75\% of bootstrapped samples, model M(st,t) outperformed M(r,t) in both rMAE and rWIS (i.e., the lower edge of the box of the box plot is at or above 0), while M(sv,v) outperformed M(r,v) in nearly 100\% of bootstrapped samples.

\subsubsection{Q2: Does joint training with real and synthetic data improve COVID-19 case forecasts relative to training with either source individually?}
\label{subsubsec:q2}

\begin{figure}[H]
  \centering
  \includegraphics[width=1\textwidth]{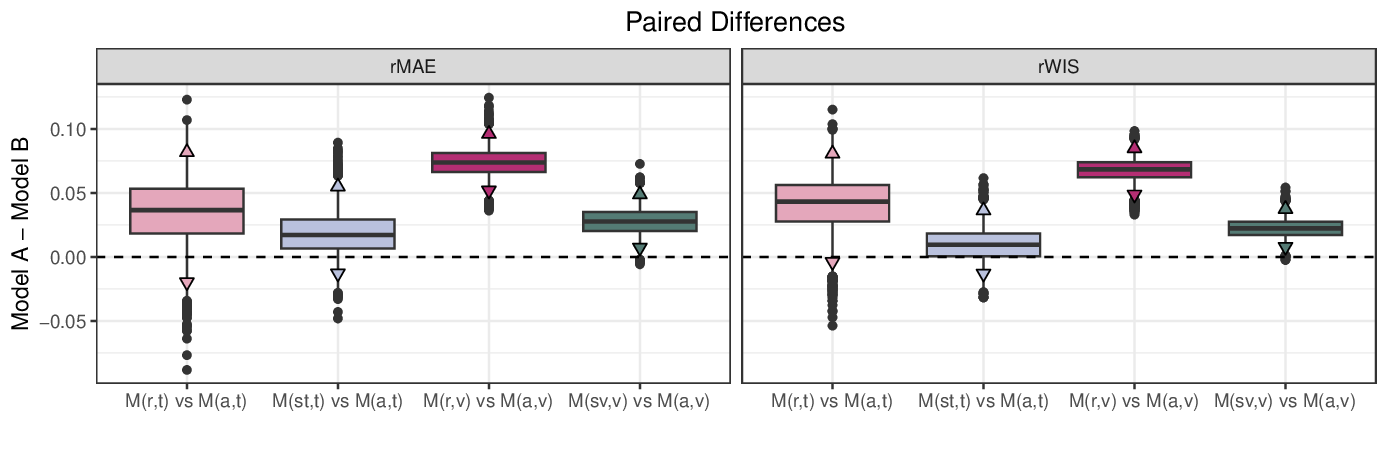}
  \caption{Distribution of 5,000 bootstrapped model differences for rMAE and rWIS. Triangles indicate the 2.5 and 97.5 percentiles. Presented results show Model A - Model B (on the x-axis it is displayed as M(A) vs M(B)).  For instance, M(r,t) vs M(a,t) in the rMAE panel presents the results of M(r,t)'s rMAE - M(a,t)'s rMAE, across all bootstrap samples. Positive numbers indicate Model B performed better than Model A. Clear evidence is shown that models trained with real and synthetic data (M(a,t) and M(a,v)), on balance, performed better than models trained with only-real or only-synthetic data.}
  \label{fig:q2}
\end{figure}

Yes, there is strong evidence that training with real \emph{and} synthetic data is better than training with either source individually.
See Figure \ref{fig:q2}.
In all comparisons, at least 75\% of bootstrapped samples resulted in model M(a,.) having better performance (rMAE or rWIS) than the analogous model trained with either real data alone or synthetic data alone. 
In all comparisons presented in Figure \ref{fig:q2}, there is no evidence that training with real and synthetic data is harmful relative to training with either source individually, making this joint training the safe choice (that is, using all available training data yielded better results than a subset).

Before moving on to question Q3, it's worth taking a moment to contemplate \emph{why} training on synthetic data alone outperformed training on real data alone (Q1) and why training on both sources outperformed either one individually (Q2).  
A simple hypothesis is related to Table \ref{tab:training_summary}: there is more synthetic training data ($\sim$9 million examples) than there is real training data ($\sim$2 million examples), and there is (by definition) more real plus synthetic training data ($\sim$20 million examples) than there is of either source individually and more training data equates to better model performance. 
That is, maybe when comparing M(i,.) to M(j,.) for two different training data types i and j, the performance can be predicted simply by knowing the amount of training data available.

\begin{figure}[H]
  \centering
  \includegraphics[width=1\textwidth]{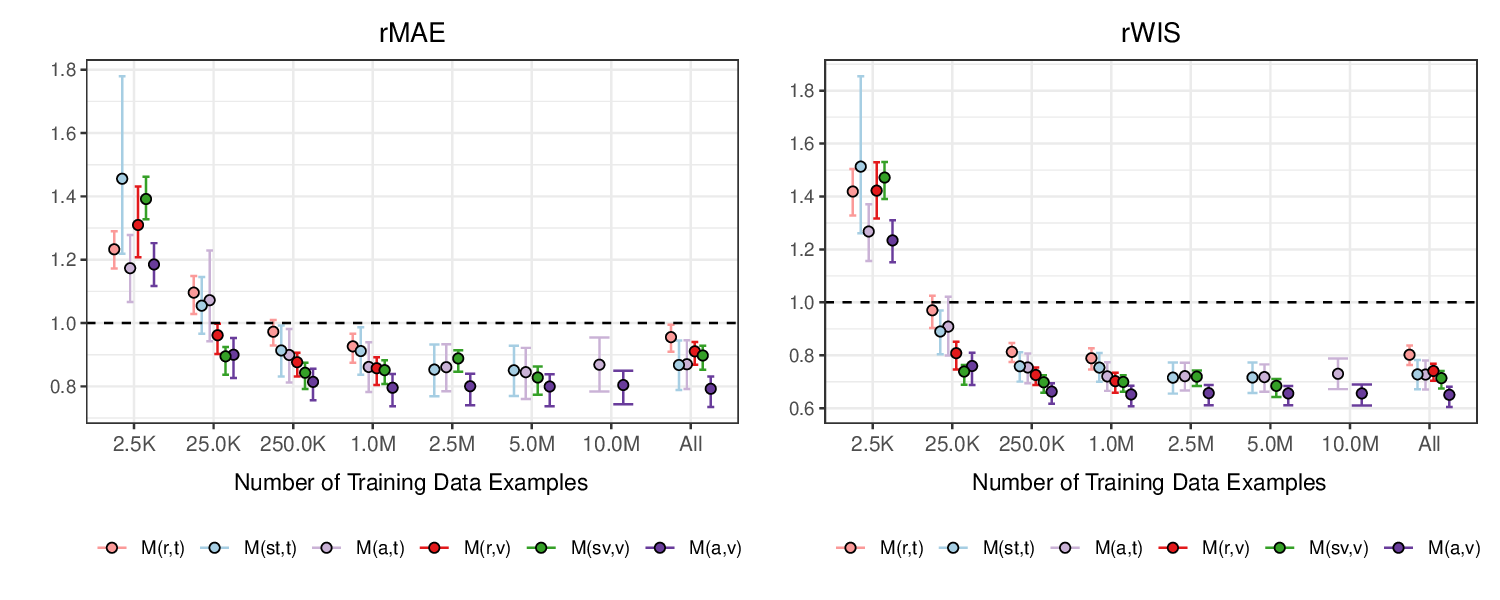}
  \caption{The relative mean absolute error (rMAE) and relative weighted interval score (rWIS) versus the number of training data examples for all models (sans M(st,v) and M(sv,t) --- the synthetic models where the training data and the input data type are mismatched). Model performance improves as training data size increases until about one million training examples are used. ``All" means the results when all available training examples for each model are used (numbers available in Table \ref{tab:training_summary}). Results presented are based on training runs with 25k model weight updates.}
  \label{fig:trainsize}
\end{figure}

To investigate this hypothesis, we retrained the models on training data sizes of $\{$2.5k, 25k, 250k, 1M, 2.5M, 5M, 10M$\}$ by taking subsets of the available training data of each type for all training data sizes less than or equal to all the available data. 
So, for clarity, M(r,.) is trained on training data sets of sizes 2.5k, 25k, 250k, 1M (all training data sizes less than the available 2.1M --- recall Table \ref{tab:training_summary}) as well as 2.1M (all the available training data).
Similarly for models M(st,.), M(sv,.) and M(a,.).
For each model/training data size, we calculate rMAE and rWIS averaged across all states, forecast dates, and forecast horizons.
If our hypothesis is correct and the amount of training data is the driving force behind forecast improvement, we would expect to see similar performance for all models when trained on the same amount of training data, holding the input data type (TC or VAC) constant.
If, however, for the same amount of training data we see some models consistently outperform other models (e.g., if M(a,.) $>$ M(r,.)), that would be evidence against our hypothesis and would suggest the amount of training data does not solely explain why some models outperform others.
Results from this exercise are shown in Figure \ref{fig:trainsize}.

\begin{figure}[H]
  \centering
  \includegraphics[width=1\textwidth]{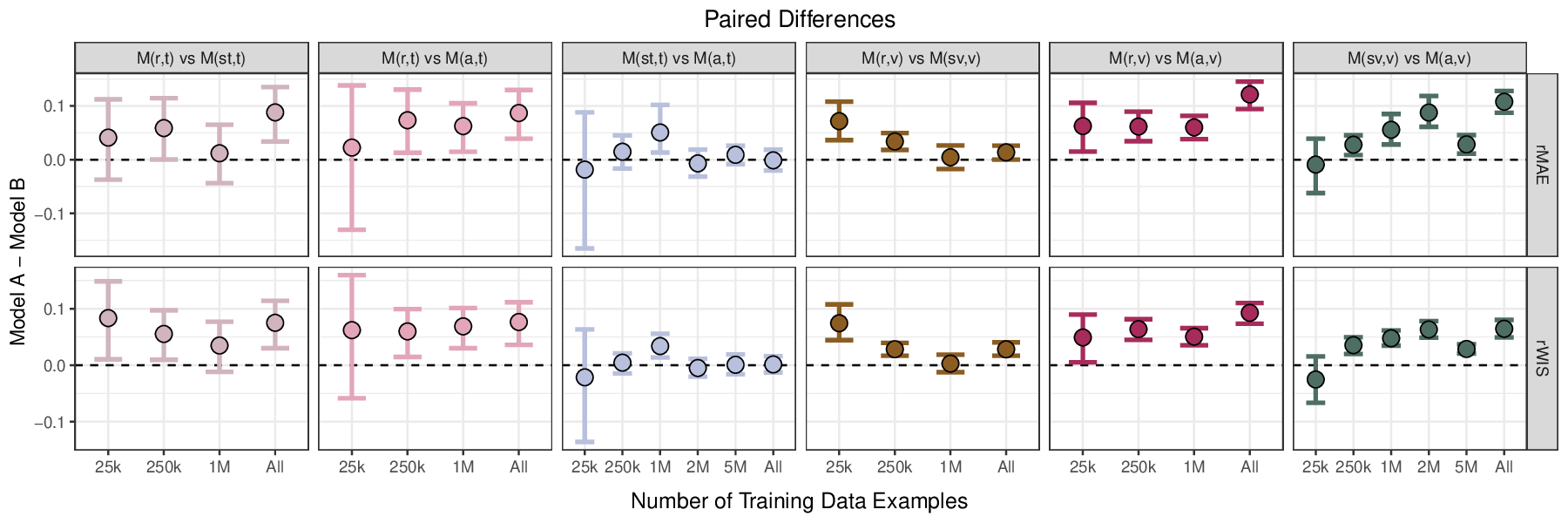}
  \caption{The 95\% confidence intervals (bars) and averages (points) in relative mean absolute error (rMAE) and relative weighted interval score (rWIS) for paired differences of models (columns) based on 5,000 bootstrap samples. Training data sizes are shown on the x-axis. ``All" means the results when all available training examples for each model are used (numbers available in Table \ref{tab:training_summary}). Results are presented as Model A - Model B, and positive values mean that Model B performed better than Model A. Results only shown for 25k or more training examples. }
  \label{fig:trainsize_paired_diff}
\end{figure}

Figure \ref{fig:trainsize} shows rMAE and rWIS for increasing amounts of training examples. 
We see that as the amount of training data increases from 2.5k examples to 1M examples, the performance of each model generally improves. 
In this regime, there is evidence that more training data correlates with better model performance.
After models are trained with 1M training examples, however, forecast performance appears to level off.
More importantly, there do appear to be systematic differences in model performance, even when holding the amount of training data constant. 
This can more clearly be seen in Figure \ref{fig:trainsize_paired_diff}.

Figure \ref{fig:trainsize_paired_diff} shows the 95\% confidence intervals for paired differences in rMAE and rWIS across 5,000 bootstrapped samples. 
While not universal, there is clear evidence shown in Figure \ref{fig:trainsize_paired_diff} that the model trained with real and synthetic training data outperforms the models trained with only real or only synthetic training data \emph{even after} holding the number of training data examples constant (2nd, 3rd, 5th, and 6th columns of Figure \ref{fig:trainsize_paired_diff}).
Furthermore, there is evidence shown that the models trained on synthetic data only outperform the models trained on real data only \emph{even after} holding the number of training data examples constant (1st and 4th columns of Figure \ref{fig:trainsize_paired_diff}).
\emph{Figure \ref{fig:trainsize_paired_diff} presents evidence against our hypothesis} that the amount of training data is the explanation for M(a,.) outperforming all other models (Q2) and M(st,t)/M(sv,v) outperforming M(r,t)/M(r,v) (Q1).
That is, something more than training data quantity is needed to explain the findings in Q1 and Q2.

Our other hypothesis for why training on real and synthetic data outperforms either source individually and why training with synthetic data only outperforms training with real data only centers around \emph{covariate shift} \cite{nair2019covariate}.
Covariate shift in supervised machine learning (ML) refers to the change in the distribution of the input examples to a ML model between the training data (i.e., non-COVID-19, real respiratory data or synthetic data) and the testing data (i.e., COVID-19 data).
Most supervised ML models assume that the training and testing data inputs (i.e., in this work, the last 20 observations of a time series) come from a common distribution. 
When there is a distributional mismatch, predictive performance can suffer.
It could be the case that COVID-19 data are better represented by synthetic data than historical non-COVID-19, respiratory data (i.e., it could be synthetic data and COVID-19 data appear more alike than non-COVID-19, real respiratory data). 

\begin{figure}[H]
  \centering
  \includegraphics[width=1\textwidth]{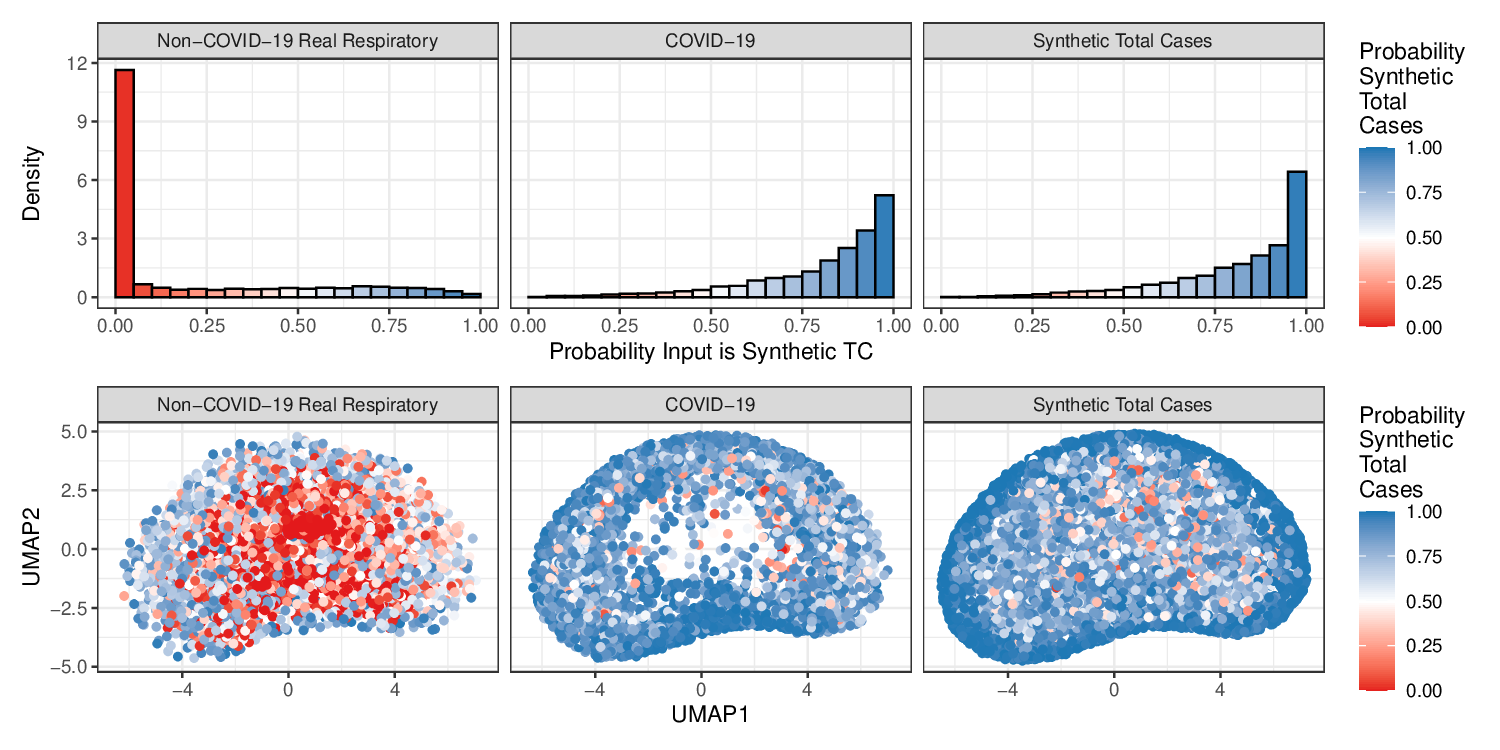}
  \caption{(top) The distribution of predicted probabilities that input data are synthetic (total cases) from a hold-out set. 8,000 hold-out examples from the non-COVID-19 real respiratory and synthetic total cases, respectively are being summarized along with 6,885 COVID-19 examples. COVID-19 data are overwhelmingly classified as synthetic data rather than non-COVID-19, real respiratory data, while the classifier correctly classifies synthetic data as synthetic data and non-COVID-19, real respiratory data as non-COVID-19, real respiratory data. (bottom) UMAP arrangement of hold-out data, colored by the predicted probability synthetic. We can see the non-COVID-19, real respiratory data occupies a different part of UMAP space that the COVID-19 data, while the COVID-19 data and synthetic data have more overlap in UMAP space, shedding light on why the classifier classifies COVID-19 data as synthetic.}
  \label{fig:umap}
\end{figure}

To investigate this possibility, we fit a gradient boosted model to perform binary classification trained on a 150k non-COVID-19, real respiratory training examples and 150k synthetic TC training examples.
The inputs to this model were the last 20 observations of a time series (mean centered and standard deviation scaled) and the labels ``Synthetic TC" or ``Real" were the output.
No COVID-19 data were used to train this classifier.
We then ran three different hold-out data sets through the classifier: 1) 8,000 instances of non-COVID-19, real respiratory data, 2) 8,000 instances of synthetic TC data,  and 3) all 6,885 instances of COVID-19 data. 
The non-COVID-19, real respiratory data and synthetic TC data sets were used to evaluate the classifiers capabilities.
The COVID-19 data were used to see if COVID-19 data better resemble non-COVID-19, real respiratory data or synthetic TC data.
If the COVID-19 data are classified as ``Synthetic TC" data, that would constitute evidence that COVID-19 data come from a distribution more similar to synthetic data than non-COVID-19, real respiratory data.
Results are shown in Figure \ref{fig:umap}. 

Overwhelmingly, COVID-19 data were classified as synthetic data rather than non-COVID-19, real respiratory data.
This is seen in the top of Figure \ref{fig:umap}.
Over 90\% of COVID-19 instances were assigned a probability over 0.5 that the instance was synthetic TC data.
The non-COVID-19, real respiratory data and the synthetic TC data were, with high probability, correctly classified as well. 
Over 90\% of synthetic TC instances from the hold-out set were correctly assigned a probability over 0.5 of being synthetic TC, while just under 80\% of the non-COVID-19, real respiratory data were assigned a probability over 0.5 of being non-COVID-19, real respiratory data. 
These results indicate that the trained classifier can discriminate between real and synthetic TC data.

The bottom of Figure \ref{fig:umap} provides a UMAP \cite{mcinnes2018umap} (Uniform Manifold Approximation and Projection) plot  --- a lower dimensional representation of the 20-dimensional inputs.
UMAP finds a low-dimensional embedding of high-dimensional data that preserves local neighborhood relationships. 
Points that are close together in the 20-dimensional space are close together in the 2-dimensional plot in Figure \ref{fig:umap} (and points that are far away remain far away).
This UMAP plot helps explain why the classifier was able to discriminate between real and synthetic data and explain why COVID-19 data was overwhelmingly classified as synthetic data.
The non-COVID-19, real respiratory data largely occupies the center of the UMAP plot.
The synthetic data largely covers the whole space, but is particularly concentrated around the outer ring. 
The COVID-19 data also largely occupies the outer ring and, notably, does not occupy the center of the UMAP shape.
Thus, there is evidence that the COVID-19 data inputs better match the distribution of inputs produced by synthetic data. 
Or, said another way, the covariate shift between non-COVID-19, real respiratory data (training) and COVID-19 data (testing) is much more pronounced than between synthetic data (training) and COVID-19 data (testing).
Because the real and synthetic data, however, occupy different regions of input space, combining them results in improved input space coverage, providing insight into why M(a,t) and M(a,v) outperformed their modeling counterparts.

\subsubsection{Q3a: Does training with synthetic VAC data improve COVID-19 forecasts relative to training with synthetic TC data?}
\begin{figure}[H]
  \centering
  \includegraphics[width= 1\textwidth]{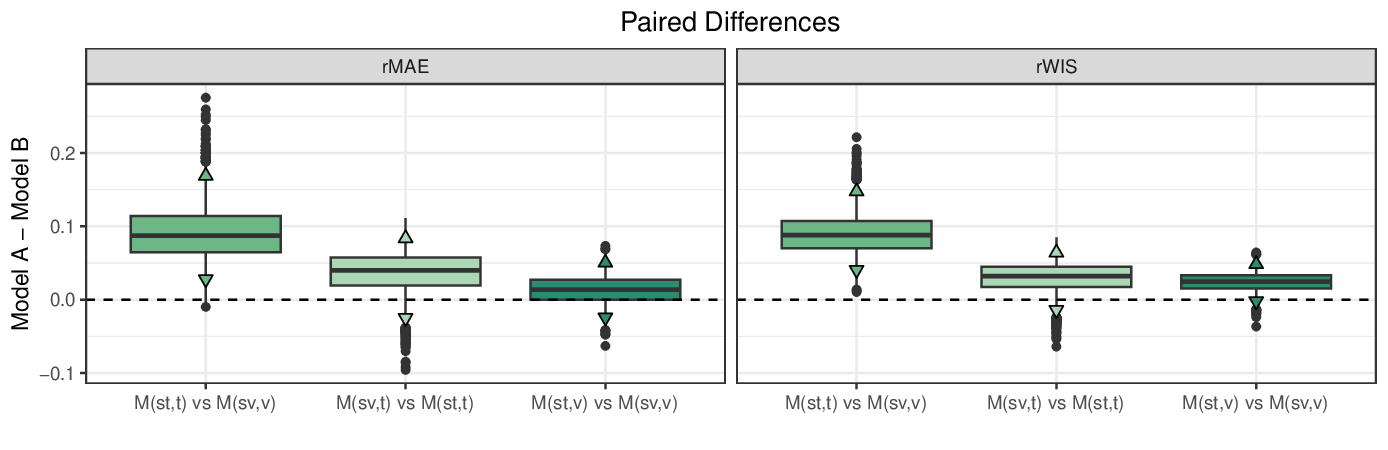}
  \caption{Distribution of 5,000 bootstrapped model differences for rMAE and rWIS. Triangles indicate the 2.5 and 97.5 percentiles. Presented results show Model A - Model B (on the x-axis it is displayed as M(A) vs M(B)). For instance, M(st,t) vs M(sv,v) in the rMAE panel presents the results of M(st,t)'s rMAE - M(sv,v)'s rMAE, across all bootstrap samples. Positive numbers indicate Model B performed better than Model A.}
  \label{fig:q3}
\end{figure}

Yes, training with synthetic VAC data resulted in better forecasts than training with synthetic TC data. 
The evidence for this is shown in the M(st,t) vs M(sv,v) comparison in Figure \ref{fig:q3}.
We see the 95\% confidence intervals do not cover 0, indicating a statistically significant improvement in forecast performance for M(sv,v) relative to M(st,t) for as measured by rMAE and rWIS.

\subsubsection{Q3b: Do models with matched training data and input data outperform models with mismatched training data and input data?}
Yes, there is clear evidence that models with matched training data and input data outperform models with mismatched training data and input data.
This can be seen in the M(sv,t) vs M(st,t) and M(st,v) vs M(sv,v) comparisons in Figure \ref{fig:q3}.
In each comparison, the model with matched training and input data (M(st,t) and M(sv,v)) outperformed their counterpart with the same input data type but different training data type in at least 75\% of bootstrapped samples. 
This makes intuitive sense.
If the model is being trained on one data type but is then being tested on a different input type, there is the potential for both covariate shift \cite{nair2019covariate} (when the distribution of inputs differs between training and testing) and concept shift \cite{iwashita2018overview} (when the relationship between inputs and outcomes differs between training and testing).

\subsubsection{Q4: Does including SARS-CoV-2 variant information improve COVID-19 case forecasts?}
\begin{figure}[H]
  \centering
  \includegraphics[width= 1\textwidth]{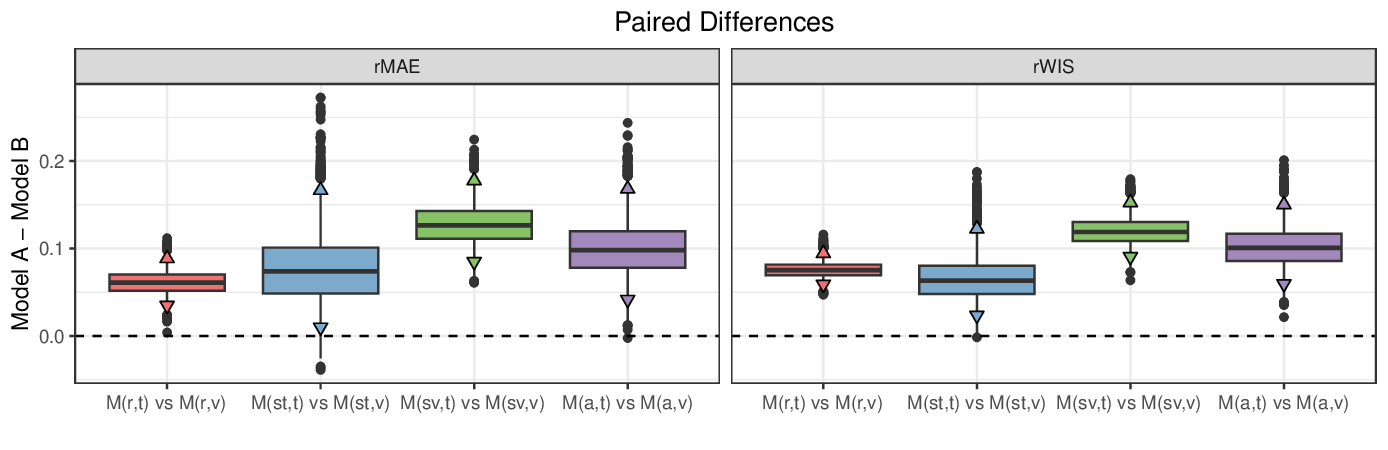}
  \caption{Distribution of 5,000 bootstrapped model differences for rMAE and rWIS. Triangles indicate the 2.5 and 97.5 percentiles. Presented results show Model A - Model B (on the x-axis it is displayed as M(A) vs M(B)). For instance, M(r,t) vs M(r,v) in the rMAE panel presents the results of M(r,t)'s rMAE - M(r,v)'s rMAE, across all bootstrap samples. Positive numbers indicate Model B performed better than Model A. Clear evidence is presented that models M(.,v) outperformed models M(.,t).}
  \label{fig:q4}
\end{figure}

Yes, there is clear evidence that forecasting variant-attributable cases directly and summing to recover a total cases forecast improves forecasts relative to forecasting total cases directly. 
See Figure \ref{fig:q4} for results.
To answer this question, we compared all pairs of M(.,t) vs M(.,v).
These comparisons hold the training data constant, and only differ in what data are passed in as inputs (total cases versus variant-attributable cases).
The 95\% confidence intervals fall above 0 for all comparisons for both metrics, indicating the models that take VACs as inputs and sum to recover total cases forecasts outperform the models that take TCs as inputs directly.
Our finding that forecasting variant-attributable cases and summing produces better forecasts than forecasting total cases mirrors findings from influenza forecasting \cite{kandula2017type,turtle2021accurate}.
We conjecture that VACs have simpler and more predictable dynamics, making them easier to forecast, than the total cases time series which is a mixture of multiple co-circulating variants. 

\begin{figure}[H]
  \centering
  \includegraphics[width= 1\textwidth]{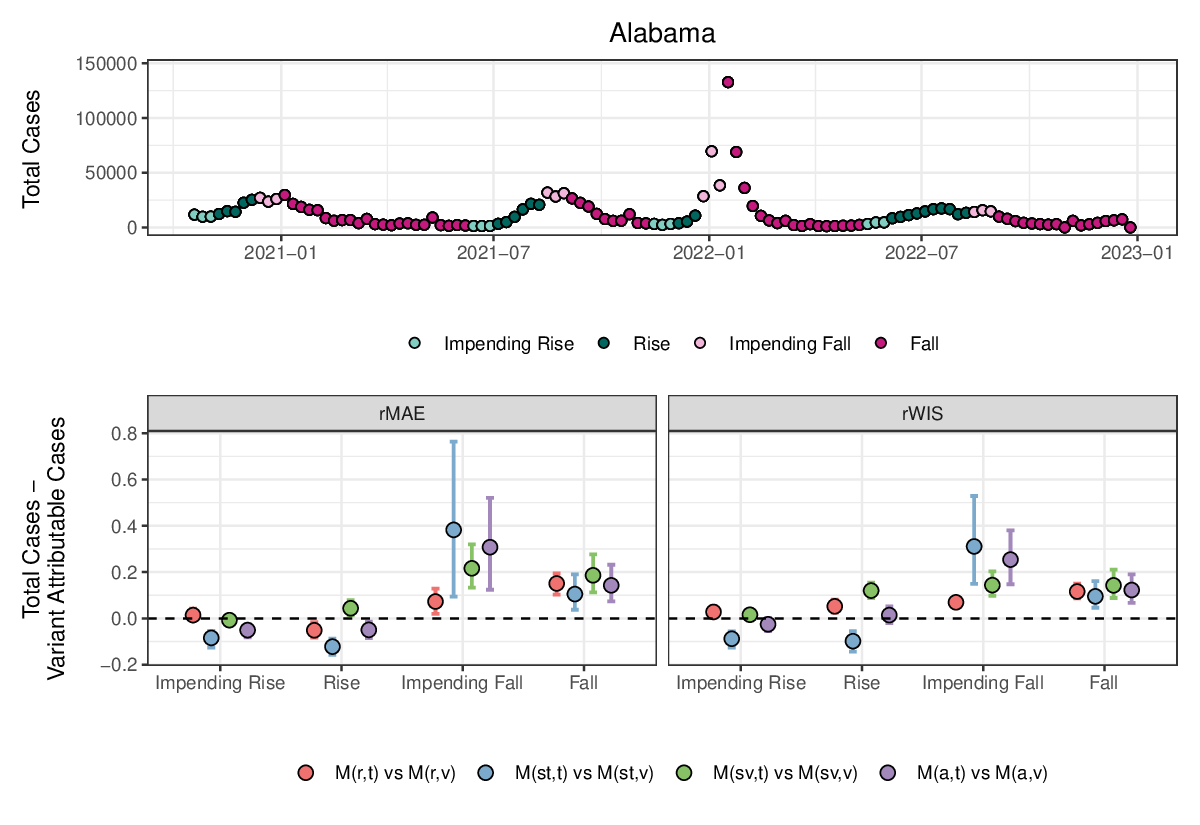}
  \caption{(top) An example (Alabama) of how the total cases time series is partitioned into four phases: impending rise, rise, impending fall, and fall. (bottom) The difference in rMAE and rWIS between models, broken down by phase for all states (not just Alabama). Points are average differences and error bars are 95\% confidence intervals based on 5,000 bootstrap samples. Values greater than 0 mean the VAC models (M(.,v)) outperformed their corresponding TC model (M(.,t)). We see that the VAC models meaningfully outperform their TC counterparts during the impending fall and fall phases of the outbreak and perform more similarly to the TC models during the impending rise and rise phases.}
  \label{fig:q4_phase}
\end{figure}

Figure \ref{fig:q4_phase} investigates during what phases of the COVID-19 pandemic models M(.,v) outperform models M(.,t). 
We partition each total cases time series into four phases: impending rise, rise, impending fall, and fall.
Those phases are illustrated in the top of Figure \ref{fig:q4_phase}.
We then compute the difference in rMAE and rWIS by phase for each pair of models. 
We see clear evidence that the M(.,v) models outperform the M(.,t) models during the impending fall and fall phases of the pandemic.
VAC and TC trained models perform more similarly to one another during the impending rise and rise phases, with TC models tending to outperform VAC models when performance is measured by MAE.
The degree to which the TC models outperform the VAC models during the rising phase(s), however, is small relative to the degree the VAC models outperform the TC models during the impending fall and fall phases.

\subsubsection{Q5: How do these forecasts compare to real-time COVID-19 case forecasts?}
\begin{figure}[H]
  \centering
  \includegraphics[width=1\textwidth]{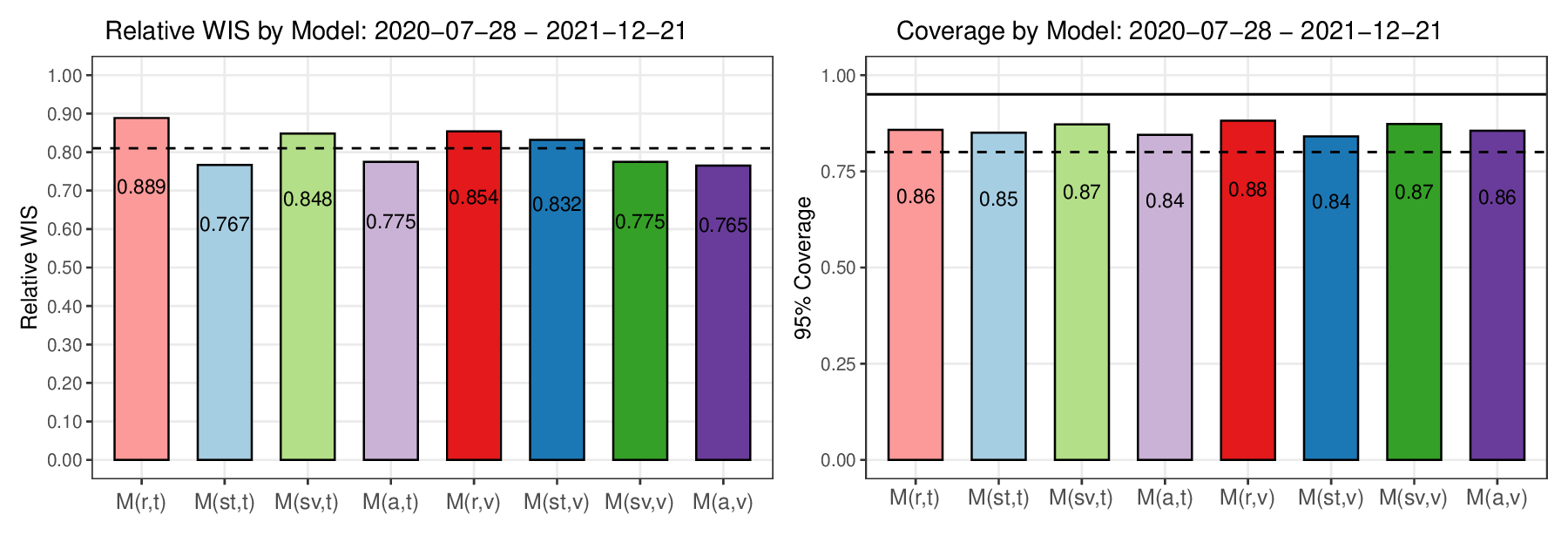}
  \caption{(left) Relative weighted interval score (rWIS) corresponding to the evaluation period of July 28th, 2020 through December 21st, 2021. Lower is better. The COVIDHub-4\_week\_ensemble had a relative WIS of 0.81 \cite{lopez2024challenges} (the horizontal dashed line). Models with rWIS below the horizontal, dashed line outperformed the COVIDHub-4\_week\_ensemble as measured by rWIS. (right) Empirical coverages for a 95\% nominal prediction interval (solid line). The COVIDHub-4\_week\_ensemble's 95\% prediction intervals had an empirical coverage of 0.8 (dashed line). All models had empirical coverages closer to nominal than that. }
  \label{fig:q5}
\end{figure}

To answer this question, we refer to the results presented in \cite{lopez2024challenges}. 
Evaluation in \cite{lopez2024challenges} included relative WIS and coverage of 1 through 4 week ahead forecasts from July 28th, 2020 through December 21st, 2021. 
See Figure \ref{fig:q5} for WIS and coverage results for all models restricted to July, 2020 through December, 2021.
All eight models outperformed a persistence (baseline) model (a feat only 7 out of 22 models accomplished in real-time).
Four models outperformed the COVIDHub-4\_week\_ensemble model with a relative WIS of 0.81 \cite{lopez2024challenges}. 
These models were the models trained on synthetic data only where the training data type matched the input data type (M(st,t) and M(sv,v)) as well as the two models trained with all the available training data (M(a,t) and M(a,v)). 
The models that did not outperform the COVIDHub-4\_week\_ensemble model were the models trained exclusively on real training data (M(r,t) and M(r,v)) and the models trained exclusively on synthetic data with a mismatch between training data type and model input type (M(sv,t) and M(st,v)).
However, even the worst performing model considered in this paper, M(r,t), would have been the 5th best performing model among the real-time forecasting models summarized in \cite{lopez2024challenges}.
Furthermore, the COVIDHub-4\_week\_ensemble had a 95\% coverage of 0.8. 
All eight models had an empirical coverage between 0.84 and 0.88, closer to nominal than did the COVIDHub-4\_week\_ensemble.
Additionally, although the VAC data input models (M(.,v)) could not have been forecast in real-time due to variant reporting delays, the TC data input models (M(.,t)) could have been forecast in real-time.
It is also worth a reminder that none of the eight models in this paper used any COVID-19 data for training; they only used data (real and/or synthetic) that would have been available on or before January 1st, 2020.

\section{Discussion}
\label{sec:discussion}

In this paper, we sought to answer two high-level questions: (1) Can synthetic data be used to train deep learning models that achieve state-of-the-art infectious disease forecasting performance? and (2) Can genetic information be used to improve infectious disease forecasting models? 
We conducted an exercise to answer these questions with an eye towards the next pandemic by limiting ourselves to only using training data available prior to January 1st, 2020 for forecasting COVID-19 cases.
We affirmatively answered both questions. 
Synthetic data was a valuable training data source, as models trained on only synthetic data provided impressive forecasting performance.
Real historical data for non-COVID-19 pathogens also proved to be a valuable training data source, though less valuable than synthetic on its own (Q1). 
Combining real and synthetic data was found to be a prudent path forward, as models trained with all available training data outperformed models trained on real or synthetic data alone (Q2).
Furthermore, forecasting variant-attributable cases directly and summing of forecasts produces demonstrably better forecasts than forecasting total cases directly (Q4).
Model M(a,v), the model that made use of all available training data and used variant information, was the best performing model. 

The deep learning models trained and used in this exercise required training once, but required no retraining.
While deep learning models can take several hours to train (recall Table \ref{tab:training_summary}), they take seconds to predict with and can scale to a large number of forecast settings.
While we trained our models once, we could have either retrained every week as new data became available, or, more commonly, performed fine-tuning on our pre-trained model with the new COVID-19 data as it became available \cite[e.g.,][]{hu2022lora}.
Thus scaling a forecasting model from, say, US states to all US counties would be straightforward.
This is in contrast to a spatial or hierarchical model that borrows information across geographies, where retraining may be required every time new data are made available and scaling up from tens to hundreds or thousands of geographies is not trivial \cite[e.g.,][]{osthus2021multiscale}.
Again, these deep learning models require adequate amounts of training data to be effective, training data that are not available in an emerging disease setting. 
Historical outbreak data of related disease and/or synthetic data proved to be useful training data.
This paper adds to the growing body of evidence that synthetic data is a valuable and under-tapped resource for infectious disease forecasting \citep{dudley2025mantis, murph2025synthetic, epifforma}, as it provides the necessary ingredient to unlock the potential of deep learning models.

There are many opportunities to push this deep learning plus synthetic data idea further.
Going back to the input/output deep learning framework, many future forecasting opportunities for improvement amount to augmenting the inputs.
Forecasting geographic locations jointly rather than one at a time is an exciting path forward (recent real-time influenza forecasting success has employed a hierarchical modeling approach that borrows information across geographies \cite{osthus2021multiscale,case2025accurate}).
This amounts to developing training examples where, say, recent observations from all US states are the input and the next H time steps for each US state is the output, allowing the deep learning model to learn (potentially) complex inter-state relationships.
To learn these relationships, however, will require suitable training data.
MutAntiGen has multi-location capabilities thus could be a suitable synthetic data generator.
Another application is jointly forecasting cases, hospitalizations, and deaths (as was done in \cite{dudley2025mantis}).
Rather than forecasting variant-attributable cases, one could consider computing population genetic summaries of the viral population as time series and providing those as inputs paired with total cases. 
This approach would require a synthetic data simulator with appropriate evolutionary biology fidelity.
In general, for this deep learning plus synthetic data idea to scale will require simulators sophisticated enough to make training data with sufficient realism, which will likely require continued simulator development.

While synthetic data is a promising resource to use, it is not a panacea. 
How to generate it requires thoughtful contemplation of the problem at hand (recall Section \ref{S-sec:obsmodel}).
For instance, for much of the COVID-19 pandemic, data was collected and disseminated in the US at daily resolution, introducing day-of-week effects. 
Such effects are learnable and exploitable, but for a deep learning model to incorporate them, it would need to be presented with training data reflecting those patterns.
We did not create any synthetic data having day-of-week effects, so we anticipate our approach might fail if applied to daily data. 
That said, to forecast  at the daily scale, we could add a day-of-week routine to Section \ref{S-sec:obsmodel}.
Similarly, to use this approach to forecast seasonal diseases, we would want to both ensure we are synthetically generating seasonal outbreaks as well to increase the context window (up to, say, C $=$ 52 or 104), capturing at least one entire period of the outbreak. 

Finally, while we conducted this retrospective forecasting study with an eye towards real-time forecasting, we acknowledge that reporting delays exist with both epidemiological and genetic data. 
Our results represent best-case scenarios with respect to reporting delays and more work is needed to bridge the real-time reporting delay gap.

\section{Acknowledgments}
\label{sec:acknowledgements}
We gratefully acknowledge all data contributors, i.e., the Authors and their Originating laboratories responsible for obtaining the specimens, and their Submitting laboratories for generating the genetic sequence and metadata and sharing via the GISAID Initiative, on which some of this research is based.
Research presented in this article was supported by the Laboratory Directed Research and Development program of Los Alamos National Laboratory under project number 20240066DR. 
Los Alamos National Laboratory is operated by Triad National Security, LLC, for the National Nuclear Security Administration of U.S. Department of Energy (Contract No. 89233218CNA000001).
The authors acknowledge the use of ChatGPT for text editing and assistance with selected coding tasks.
The authors retain full responsibility for the manuscript's content.
This article approved for unlimited release (LA-UR-26-22310).

\newpage
\bibliographystyle{apalike}
\bibliography{bibliography}

@article{murph2025synthetic,
  title={Synthetic method of analogues for emerging infectious disease forecasting},
  author={Murph, Alexander C and Gibson, G Casey and Amona, Elizabeth B and Beesley, Lauren J and Castro, Lauren A and Del Valle, Sara Y and Osthus, Dave},
  journal={PLOS Computational Biology},
  volume={21},
  number={6},
  pages={e1013203},
  year={2025},
  publisher={Public Library of Science San Francisco, CA USA}
}

@article{du2023incorporating,
  title={{Incorporating variant frequencies data into short-term forecasting for COVID-19 cases and deaths in the USA: a deep learning approach}},
  author={Du, Hongru and Dong, Ensheng and Badr, Hamada S and Petrone, Mary E and Grubaugh, Nathan D and Gardner, Lauren M},
  journal={{eBioMedicine}},
  volume={89},
  year={2023},
  publisher={Elsevier}
}

@article{osthus2021multiscale,
  title={Multiscale influenza forecasting},
  author={Osthus, Dave and Moran, Kelly R},
  journal={{Nature Communications}},
  volume={12},
  number={1},
  pages={2991},
  year={2021},
  publisher={Nature Publishing Group UK London}
}

@article{brooks2015flexible,
  title={{Flexible modeling of epidemics with an empirical Bayes framework}},
  author={Brooks, Logan C and Farrow, David C and Hyun, Sangwon and Tibshirani, Ryan J and Rosenfeld, Roni},
  journal={{PLoS Computational Biology}},
  volume={11},
  number={8},
  pages={e1004382},
  year={2015},
  publisher={Public Library of Science San Francisco, CA USA}
}

@article{ray2025flusion,
  title={{Flusion: Integrating multiple data sources for accurate influenza predictions}},
  author={Ray, Evan L and Wang, Yijin and Wolfinger, Russell D and Reich, Nicholas G},
  journal={Epidemics},
  volume={50},
  pages={100810},
  year={2025},
  publisher={Elsevier}
}

@article{dong2020interactive,
  title={{An interactive web-based dashboard to track COVID-19 in real time}},
  author={Dong, Ensheng and Du, Hongru and Gardner, Lauren},
  journal={{The Lancet Infectious Diseases}},
  volume={20},
  number={5},
  pages={533--534},
  year={2020},
  publisher={Elsevier}
}

@article{lopez2024challenges,
  title={{Challenges of COVID-19 Case Forecasting in the US, 2020--2021}},
  author={Lopez, Velma K and Cramer, Estee Y and Pagano, Robert and Drake, John M and O’Dea, Eamon B and Adee, Madeline and Ayer, Turgay and Chhatwal, Jagpreet and Dalgic, Ozden O and Ladd, Mary A and others},
  journal={{PLoS Computational Biology}},
  volume={20},
  number={5},
  pages={e1011200},
  year={2024},
  publisher={Public Library of Science San Francisco, CA USA}
}

@article{beesley2022addressing,
  title={{Addressing delayed case reporting in infectious disease forecast modeling}},
  author={Beesley, Lauren J and Osthus, Dave and Del Valle, Sara Y},
  journal={{PLoS Computational Biology}},
  volume={18},
  number={6},
  pages={e1010115},
  year={2022},
  publisher={Public Library of Science San Francisco, CA USA}
}

@article{shadbolt2022challenges,
  title={{The challenges of data in future pandemics}},
  author={Shadbolt, Nigel and Brett, Alys and Chen, Min and Marion, Glenn and McKendrick, Iain J and Panovska-Griffiths, Jasmina and Pellis, Lorenzo and Reeve, Richard and Swallow, Ben},
  journal={Epidemics},
  volume={40},
  pages={100612},
  year={2022},
  publisher={Elsevier}
}

@article{ioannidis2022forecasting,
  title={{Forecasting for COVID-19 has failed}},
  author={Ioannidis, John PA and Cripps, Sally and Tanner, Martin A},
  journal={{International Journal of Forecasting}},
  volume={38},
  number={2},
  pages={423--438},
  year={2022},
  publisher={Elsevier}
}

@article{cramer2022united,
  title={{The United States COVID-19 forecast hub dataset}},
  author={Cramer, Estee Y and Huang, Yuxin and Wang, Yijin and Ray, Evan L and Cornell, Matthew and Bracher, Johannes and Brennen, Andrea and Rivadeneira, Alvaro J Castro and Gerding, Aaron and House, Katie and Jayawardena, Dasuni and Kanji, Abdul Hannan and Khandelwal, Ayush and Le, Khoa and Mody, Vidhi and Mody, Vrushti and Niemi, Jarad and Stark, Ariane and Shah, Apurv and Wattanchit, Nutcha and Zorn, Martha W amd Reich, Nicholas G and US COVID-19 Forecast Hub Consortium},
  journal={{Scientific Data}},
  volume={9},
  number={1},
  pages={462},
  year={2022},
  publisher={Nature Publishing Group UK London}
}

@article{sherratt2023predictive,
  title={{Predictive performance of multi-model ensemble forecasts of COVID-19 across European nations}},
  author = {Sherratt, Katharine and Gruson, Hugo and Grah, Rok and Johnson, Helen and Niehus, Rene and
Prasse, Bastian and Sandmann, Frank and Deuschel, Jannik and Wolffram, Daniel and
Abbott, Sam and Ullrich, Alexander and Gibson, Graham and Ray, Evan L and Reich, Nicholas G and
Sheldon, Daniel and Wang, Yijin and Wattanachit, Nutcha and Wang, Lijing and Trnka, Jan and
Obozinski, Guillaume and Sun, Tao and Thanou, Dorina and Pottier, Loic and
Krymova, Ekaterina and Meinke, Jan H and Barbarossa, Maria Vittoria and
Leithauser, Neele and Mohring, Jan and Schneider, Johanna and Wlazlo, Jaroslaw and
Fuhrmann, Jan and Lange, Berit and Rodiah, Isti and Baccam, Prasith and Gurung, Heidi and
Stage, Steven and Suchoski, Bradley and Budzinski, Jozef and Walraven, Robert and
Villanueva, Inmaculada and Tucek, Vit and Smid, Martin and Zajicek, Milan and
Perez Alvarez, Cesar and Reina, Borja and Bosse, Nikos I and Meakin, Sophie R and
Castro, Lauren and Fairchild, Geoffrey and Michaud, Isaac and Osthus, Dave and
Alaimo Di Loro, Pierfrancesco and Maruotti, Antonello and Eclerova, Veronika and
Kraus, Andrea and Kraus, David and Pribylova, Lenka and Dimitris, Bertsimas and
Li, Michael Lingzhi and Saksham, Soni and Dehning, Jonas and Mohr, Sebastian and
Priesemann, Viola and Redlarski, Grzegorz and Bejar, Benjamin and Ardenghi, Giovanni and
Parolini, Nicola and Ziarelli, Giovanni and Bock, Wolfgang and Heyder, Stefan and
Hotz, Thomas and Singh, David E and Guzman-Merino, Miguel and Aznarte, Jose L and
Morina, David and Alonso, Sergio and Alvarez, Enric and Lopez, Daniel and Prats, Clara and
Burgard, Jan Pablo and Rodloff, Arne and Zimmermann, Tom and Kuhlmann, Alexander and
Zibert, Janez and Pennoni, Fulvia and Divino, Fabio and Catala, Marti and
Lovison, Gianfranco and Giudici, Paolo and Tarantino, Barbara and Bartolucci, Francesco and
Jona Lasinio, Giovanna and Mingione, Marco and Farcomeni, Alessio and
Srivastava, Ajitesh and Montero-Manso, Pablo and Adiga, Aniruddha and Hurt, Benjamin and
Lewis, Bryan and Marathe, Madhav and Porebski, Przemyslaw and
Venkatramanan, Srinivasan and Bartczuk, Rafal P and Dreger, Filip and Gambin, Anna and
Gogolewski, Krzysztof and Gruziel-Slomka, Magdalena and Krupa, Bartosz and
Moszyński, Antoni and Niedzielewski, Karol and Nowosielski, Jedrzej and
Radwan, Maciej and Rakowski, Franciszek and Semeniuk, Marcin and Szczurek, Ewa and
Zielinski, Jakub and Kisielewski, Jan and Pabjan, Barbara and Holger, Kirsten and
Kheifetz, Yuri and Scholz, Markus and Biecek, Przemyslaw and Bodych, Marcin and
Filinski, Maciej and Idzikowski, Radoslaw and Krueger, Tyll and Ozanski, Tomasz and
Bracher, Johannes and Funk, Sebastian}, 
  journal={{eLife}},
  volume={12},
  pages={e81916},
  year={2023},
  publisher={eLife Sciences Publications Limited}
}

@article{korber2020tracking,
  title={{Tracking changes in SARS-CoV-2 spike: evidence that D614G increases infectivity of the COVID-19 virus}},
  author={Korber, Bette and Fischer, Will M. and Gnanakaran, Sandrasegaram and Yoon, Hyejin and Theiler, James and Abfalterer, Werner and Hengartner, Nick and Giorgi, Elena E. and Bhattacharya, Tanmoy and Foley, Brian and Hastie, Kathryn M. and Parker, Matthew D. and Partridge, David G. and Evans, Cariad M. and Freeman, Timothy M. and de Silva, Thushan I. and McDanal, Charlene and Perez, Lautaro G. and Tang, Haili and Moon-Walker, Alex and Whelan, Sean P. and LaBranche, Celia C. and Saphire, Erica O. and Montefiori, David C.},
  journal={Cell},
  volume={182},
  number={4},
  pages={812--827},
  year={2020},
  publisher={Elsevier}
}

@book{hyndman2018forecasting,
  title={{Forecasting: Principles and Practice}},
  author={Hyndman, Rob J and Athanasopoulos, George},
  year={2018},
  publisher={{OTexts}}
}

@article{brauer2008compartmental,
  title={{Compartmental models in epidemiology}},
  author={Brauer, Fred},
  journal={{Mathematical Epidemiology}},
  pages={19--79},
  year={2008},
  publisher={Springer}
}

@article{tracy2018agent,
  title={Agent-based modeling in public health: current applications and future directions},
  author={Tracy, Melissa and Cerd{\'a}, Magdalena and Keyes, Katherine M},
  journal={{Annual Review of Public Health}},
  volume={39},
  number={1},
  pages={77--94},
  year={2018},
  publisher={Annual Reviews}
}

@article{featherstone2022epidemiological,
  title={Epidemiological inference from pathogen genomes: a review of phylodynamic models and applications},
  author={Featherstone, Leo A and Zhang, Joshua M and Vaughan, Timothy G and Duchene, Sebastian},
  journal={{Virus Evolution}},
  volume={8},
  number={1},
  pages={veac045},
  year={2022},
  publisher={Oxford University Press UK}
}

@misc{ncsl2023,
  author       = {{Kolman, Shannon}},
  title        = {{Disease Forecasting Tools Can Support Policymaking During Epidemics}},
  year         = {2023},
  url          = {https://www.ncsl.org/health/disease-forecasting-tools-can-support-policymaking-during-epidemics},
  note         = {Accessed: 2025-07-13, \url{https://www.ncsl.org/health/disease-forecasting-tools-can-support-policymaking-during-epidemics}}
}

@article{lyu2023human,
  title={{Human behavior in the time of COVID-19: Learning from big data}},
  author={Lyu, Hanjia and Imtiaz, Arsal and Zhao, Yufei and Luo, Jiebo},
  journal={{Frontiers in Big Data}},
  volume={6},
  pages={1099182},
  year={2023},
  publisher={Frontiers Media SA}
}

@article{kapitsinis2020underlying,
  title={{The underlying factors of the COVID-19 spatially uneven spread. Initial evidence from regions in nine EU countries}},
  author={Kapitsinis, Nikos},
  journal={{Regional Science Policy \& Practice}},
  volume={12},
  number={6},
  pages={1027--1046},
  year={2020},
  publisher={Elsevier}
}

@article{ling2022challenges,
  title={{Challenges and opportunities for global genomic surveillance strategies in the COVID-19 era}},
  author={Ling-Hu, Ted and Rios-Guzman, Estefany and Lorenzo-Redondo, Ramon and Ozer, Egon A and Hultquist, Judd F},
  journal={{Viruses}},
  volume={14},
  number={11},
  pages={2532},
  year={2022},
  publisher={MDPI}
}

@article{mcinnes2018umap,
  title={{UMAP: Uniform manifold approximation and projection for dimension reduction}},
  author={McInnes, Leland and Healy, John and Melville, James},
  journal={{arXiv preprint arXiv:1802.03426}},
  year={2018}
}

@article{cramer2022evaluation,
  title={{Evaluation of individual and ensemble probabilistic forecasts of COVID-19 mortality in the United States}},
  author={Cramer, Estee Y and others},
  journal={{Proceedings of the National Academy of Sciences}},
  volume={119},
  number={15},
  pages={e2113561119},
  year={2022},
  publisher={National Academy of Sciences}
}

@article{Badu2023,
	author = {Badr, Hamada S. and Zaitchik, Benjamin F. and Kerr, Gaige H. and Nguyen, Nhat-Lan H. and Chen, Yen-Ting and Hinson, Patrick and Colston, Josh M. and Kosek, Margaret N. and Dong, Ensheng and Du, Hongru and Marshall, Maximilian and Nixon, Kristen and Mohegh, Arash and Goldberg, Daniel L. and Anenberg, Susan C. and Gardner, Lauren M.},
	date = {2023/06/07},
	doi = {10.1038/s41597-023-02276-y},
	id = {Badr2023},
	isbn = {2052-4463},
	journal = {Scientific Data},
	number = {1},
	pages = {367},
	title = {Unified real-time environmental-epidemiological data for multiscale modeling of the COVID-19 pandemic},
	url = {https://doi.org/10.1038/s41597-023-02276-y},
	volume = {10},
	year = {2023}}

@article{dudley2025mantis,
  title={Mantis: A simulation-grounded foundation model for disease forecasting},
  author={Dudley, Carson and Magdaleno, Reiden and Harding, Christopher and Sharma, Ananya and Martin, Emily and Eisenberg, Marisa},
  journal={arXiv preprint arXiv:2508.12260},
  year={2025}
}

@article{bracher2021evaluating,
  title={Evaluating epidemic forecasts in an interval format},
  author={Bracher, Johannes and Ray, Evan L and Gneiting, Tilmann and Reich, Nicholas G},
  journal={{PLoS Computational Biology}},
  volume={17},
  number={2},
  pages={e1008618},
  year={2021},
  publisher={Public Library of Science San Francisco, CA USA}
}

@inproceedings{nair2019covariate,
  title={{Covariate shift: A review and analysis on classifiers}},
  author={Nair, Nimisha G and Satpathy, Pallavi and Christopher, Jabez and others},
  booktitle={{2019 Global Conference for Advancement in Technology (GCAT)}},
  pages={1--6},
  year={2019},
  organization={IEEE}
}

@article{kandula2017type,
  title={Type-and subtype-specific influenza forecast},
  author={Kandula, Sasikiran and Yang, Wan and Shaman, Jeffrey},
  journal={{American Journal of Epidemiology}},
  volume={185},
  number={5},
  pages={395--402},
  year={2017},
  publisher={Oxford University Press}
}

@article{turtle2021accurate,
  title={Accurate influenza forecasts using type-specific incidence data for small geographic units},
  author={Turtle, James and Riley, Pete and Ben-Nun, Michal and Riley, Steven},
  journal={{PLoS Computational Biology}},
  volume={17},
  number={7},
  pages={e1009230},
  year={2021},
  publisher={Public Library of Science San Francisco, CA USA}
}

@article{mcgowan2019collaborative,
  title={Collaborative efforts to forecast seasonal influenza in the United States, 2015--2016},
  author={McGowan, Craig J and Biggerstaff, Matthew and Johansson, Michael and Apfeldorf, Karyn M and Ben-Nun, Michal and Brooks, Logan and Convertino, Matteo and Erraguntla, Madhav and Farrow, David C and Freeze, John and 
          Ghosh, Saurav and 
          Hyun, Sangwon and 
          Kandula, Sasikiran and 
          Lega, Joceline and 
          Liu, Yang and 
          Michaud, Nicholas and 
          Morita, Haruka and 
          Niemi, Jarad and 
          Ramakrishnan, Naren and 
          Ray, Evan L. and 
          Reich, Nicholas G. and
          Riley, Pete and 
          Shaman, Jeffrey and 
          Tibshirani, Ryan and 
          Vespignani, Alessandro and 
          Zhang, Qian and 
          Reed, Carrie and 
          {The Influenza Forecasting Working Group}},
  journal={{Scientific Reports}},
  volume={9},
  number={1},
  pages={683},
  year={2019},
  publisher={Nature Publishing Group UK London}
}

@article{roster2022forecasting,
  title={Forecasting new diseases in low-data settings using transfer learning},
  author={Roster, Kirstin and Connaughton, Colm and Rodrigues, Francisco A},
  journal={Chaos, Solitons \& Fractals},
  volume={161},
  pages={112306},
  year={2022},
  publisher={Elsevier}
}

@misc{project_tycho,
  author       = {{University of Pittsburgh}},
  title        = {Project Tycho},
  year         = {2026},
  howpublished = {\url{https://www.tycho.pitt.edu/data/}},
  note         = {Accessed 2026-02-03}
}

@misc{jhustopscovid,
  author       = {{Johns Hopkins University}},
  title        = {{Johns Hopkins COVID-19 data hub ends after three years}},
  year         = {2023},
  howpublished = {\url{https://hub.jhu.edu/2023/03/10/coronavirus-resource-center-data-hub-ends/}},
  note         = {Accessed 2026-02-04}
}

@book{efron1994introduction,
  title={An introduction to the bootstrap},
  author={Efron, Bradley and Tibshirani, Robert J},
  year={1994},
  publisher={Chapman and Hall/CRC}
}

@article{case2025accurate,
  title={An accurate hierarchical model to forecast diverse seasonal infectious diseases},
  author={Case, BKM and Salcedo, Mariah Victoria and Fox, Spencer J},
  journal={medRxiv},
  pages={2025--03},
  year={2025},
  publisher={Cold Spring Harbor Laboratory Press}
}

@article{bonabeau2002,
	author = {Bonabeau, Eric},
	journal = {Proc Natl Acad Sci U S A},
	month = {May},
	number = {Suppl 3},
	pages = {7280--7287},
	title = {Agent-based modeling: methods and techniques for simulating human systems.},
	volume = {99 Suppl 3},
	year = {2002}}

@article{castro2020,
	author = {Castro, Lauren A. AND Bedford, Trevor AND Ancel Meyers, Lauren},
	journal = {PLOS Computational Biology},
	month = {02},
	number = {2},
	pages = {1-23},
	title = {Early prediction of antigenic transitions for influenza A/H3N2},
	volume = {16},
	year = {2020}}

@article{bedford2012,
	author = {Bedford, Trevor and Rambaut, Andrew and Pascual, Mercedes},
	journal = {BMC Biology},
	number = {1},
	pages = {38},
	title = {Canalization of the evolutionary trajectory of the human influenza virus},
	volume = {10},
	year = {2012}}

@article{koelle2015,
	author = {Koelle, Katia and Rasmussen, David A},
	journal = {eLife},
	month = {sep},
	pages = {e07361},
	title = {The effects of a deleterious mutation load on patterns of influenza A/H3N2's antigenic evolution in humans},
	volume = 4,
	year = 2015}

@book{gilbert2019,
  title={Agent-based models},
  author={Gilbert, Nigel},
  year={2019},
  publisher={Sage Publications}
}

@article{vaswani2017attention,
  title={Attention is all you need},
  author={Vaswani, Ashish and Shazeer, Noam and Parmar, Niki and Uszkoreit, Jakob and Jones, Llion and Gomez, Aidan N and Kaiser, {\L}ukasz and Polosukhin, Illia},
  journal={Advances in neural information processing systems},
  volume={30},
  year={2017}
}

@article{lecun2015deep,
  title={Deep learning},
  author={LeCun, Yann and Bengio, Yoshua and Hinton, Geoffrey},
  journal={nature},
  volume={521},
  number={7553},
  pages={436--444},
  year={2015},
  publisher={Nature Publishing Group UK London}
}

@article{hornik1989multilayer,
  title={Multilayer feedforward networks are universal approximators},
  author={Hornik, Kurt and Stinchcombe, Maxwell and White, Halbert},
  journal={Neural networks},
  volume={2},
  number={5},
  pages={359--366},
  year={1989},
  publisher={Elsevier}
}

@article{devroye2006nonuniform,
  title={Nonuniform random variate generation},
  author={Devroye, Luc},
  journal={Handbooks in operations research and management science},
  volume={13},
  pages={83--121},
  year={2006},
  publisher={Elsevier}
}

@article{mckay1979,
 ISSN = {00401706},
 URL = {http://www.jstor.org/stable/1268522},
 author = {M. D. McKay and R. J. Beckman and W. J. Conover},
 journal = {Technometrics},
 number = {2},
 pages = {239--245},
 publisher = {[Taylor & Francis, Ltd., American Statistical Association, American Society for Quality]},
 title = {A Comparison of Three Methods for Selecting Values of Input Variables in the Analysis of Output from a Computer Code},
 urldate = {2026-02-12},
 volume = {21},
 year = {1979}
}

@article{drake1999,
  title={Mutation rates among RNA viruses},
  author={Drake, John W and Holland, John J},
  journal={Proceedings of the National Academy of Sciences},
  volume={96},
  number={24},
  pages={13910--13913},
  year={1999},
  publisher={The National Academy of Sciences}
}

@article{holmes2009,
   author = "Holmes, Edward C.",
   title = "The Evolutionary Genetics of Emerging Viruses", 
   journal= "Annual Review of Ecology, Evolution, and Systematics",
   year = "2009",
   volume = "40",
   number = "Volume 40, 2009",
   pages = "353-372",
   doi = "https://doi.org/10.1146/annurev.ecolsys.110308.120248",
   url = "https://www.annualreviews.org/content/journals/10.1146/annurev.ecolsys.110308.120248",
   publisher = "Annual Reviews",
   issn = "1545-2069",
   type = "Journal Article",
  }

@article{korber2025,
title = {Real-time monitoring of {SARS-CoV-2} evolution during the {COVID-19} pandemic},
journal = {Cell Host \& Microbe},
volume = {33},
number = {11},
pages = {1802-1806},
year = {2025},
issn = {1931-3128},
doi = {https://doi.org/10.1016/j.chom.2025.10.013},
url = {https://www.sciencedirect.com/science/article/pii/S1931312825004275},
author = {Bette Korber and Will Fischer and James Theiler}
}

@unpublished{epifforma,
  title        = {{Beyond equal weights: A disease-agnostic approach to ensemble learning for infectious disease forecasting}},
  author       = {Murph, Alexander C. and Beesley, Lauren J. and Gibson, G. Casey and Castro, Lauren A. and Del Valle, Sara Y. and Osthus, Dave},
  note         = {{Manuscript under review at Nature Communications}},
  year         = {2026}
}

@article{shu2017gisaid,
  title={{GISAID: Global initiative on sharing all influenza data--from vision to reality}},
  author={Shu, Yuelong and McCauley, John},
  journal={Eurosurveillance},
  volume={22},
  number={13},
  pages={30494},
  year={2017}
}

@article{o2021assignment,
  title        = {Assignment of epidemiological lineages in an emerging pandemic using the pangolin tool},
  author       = {O'Toole, {\'A}ine and Scher, Emily and Underwood, Anthony and Jackson, Ben and Hill, Verity and McCrone, John T. and Colquhoun, Rachel and Ruis, Chris and Abu-Dahab, Khalil and Taylor, Ben and Yeats, Corin and Du Plessis, Louis and Maloney, Daniel and Medd, Nathan and Attwood, Stephen W. and Aanensen, David M. and Holmes, Edward C. and Pybus, Oliver G. and Rambaut, Andrew},
  journal={Virus evolution},
  volume={7},
  number={2},
  pages={veab064},
  year={2021},
  publisher={Oxford University Press UK}
}

@article{beesley2023sars,
  title={{SARS-CoV-2 variant transition dynamics are associated with vaccination rates, number of co-circulating variants, and convalescent immunity}},
  author={Beesley, Lauren J and Moran, Kelly R and Wagh, Kshitij and Castro, Lauren A and Theiler, James and Yoon, Hyejin and Fischer, Will and Hengartner, Nick W and Korber, Bette and Del Valle, Sara Y},
  journal={{EBioMedicine}},
  volume={91},
  year={2023},
  publisher={Elsevier}
}

@article{brito2022global,
  title={{Global disparities in SARS-CoV-2 genomic surveillance}},
  author={Brito, Anderson F. and Semenova, Elizaveta and Dudas, Gytis and Hassler, Gabriel W. and Kalinich, Chaney C. and Kraemer, Moritz U. G. and Ho, Joses and Tegally, Houriiyah and Githinji, George and Agoti, Charles N. and Matkin, Lucy E. and Whittaker, Charles and Bulgarian SARS-CoV-2 sequencing group and Communicable Diseases Genomics Network (Australia and New Zealand) and COVID-19 Impact Project and Danish Covid-19 Genome Consortium and Fiocruz COVID-19 Genomic Surveillance Network and GISAID core curation team and Network for Genomic Surveillance in South Africa (NGS-SA) and Swiss SARS-CoV-2 Sequencing Consortium and Howden, Benjamin P. and Sintchenko, Vitali and Zuckerman, Neta S. and Mor, Orna and Blankenship, Heather M. and de Oliveira, Tulio and Lin, Raymond T. P. and Mendonça Siqueira, Marilda and Resende, Paola Cristina and R. Vasconcelos, Ana Tereza and Spilki, Fernando R. and Santana Aguiar, Renato and Alexiev, Ivailo and Ivanov, Ivan N. and Philipova, Ivva and Carrington, Christine V. F. and S. D. Sahadeo, Nikita and Branda, Ben and Gurry, Céline and Maurer-Stroh, Sebastian and Naidoo, Dhamari and von Eije, Karin J. and Perkins, Mark D. and van Kerkhove, Maria and Hill, Sarah C. and Sabino, Ester C. and Pybus, Oliver G. and Dye, Christopher and Bhatt, Samir and Flaxman, Seth and Suchard, Marc A. and Grubaugh, Nathan D. and Baele, Guy and Faria, Nuno R.},
  journal={{Nature Communications}},
  volume={13},
  number={1},
  pages={7003},
  year={2022},
  publisher={Nature Publishing Group UK London}
}

@misc{who2020don229,
  author       = {{World Health Organization}},
  title        = {{Pneumonia of unknown cause -- China}},
  howpublished = {\url{https://www.who.int/emergencies/disease-outbreak-news/item/2020-DON229}},
  note         = {Accessed: 2026-03-04},
  year         = {2020},
  month        = jan,
  day          = {5},
  note         = {Disease Outbreak News. Accessed 4 March 2026}
}

@article{gneiting2007probabilistic,
  title={Probabilistic forecasts, calibration and sharpness},
  author={Gneiting, Tilmann and Balabdaoui, Fadoua and Raftery, Adrian E},
  journal={{Journal of the Royal Statistical Society Series B: Statistical Methodology}},
  volume={69},
  number={2},
  pages={243--268},
  year={2007},
  publisher={Oxford University Press}
}

@article{iwashita2018overview,
  title={An overview on concept drift learning},
  author={Iwashita, Adriana Sayuri and Papa, Joao Paulo},
  journal={IEEE access},
  volume={7},
  pages={1532--1547},
  year={2018},
  publisher={IEEE}
}

@article{hu2022lora,
  title={Lora: Low-rank adaptation of large language models.},
  author={Hu, Edward J and Shen, Yelong and Wallis, Phillip and Allen-Zhu, Zeyuan and Li, Yuanzhi and Wang, Shean and Wang, Liang and Chen, Weizhu},
  journal={Iclr},
  volume={1},
  number={2},
  pages={3},
  year={2022}
}

\clearpage

\begin{center}
{\LARGE\bfseries Supplemental Material to ``Leveraging Sequence and Synthetic Data to Improve Epidemic Forecasting"\par}
\end{center}


\setcounter{section}{0}
\renewcommand{\thesection}{S\arabic{section}}

\setcounter{subsection}{0}
\renewcommand{\thesubsection}{S\arabic{section}.\arabic{subsection}}

\renewcommand{\theHsection}{S\arabic{section}}
\renewcommand{\theHsubsection}{S\arabic{section}.\arabic{subsection}}
\renewcommand{\theHsubsubsection}{S\arabic{section}.\arabic{subsection}.\arabic{subsubsection}}

\setcounter{figure}{0}
\renewcommand{\thefigure}{S\arabic{figure}}
\renewcommand{\theHfigure}{S\arabic{figure}}

\setcounter{table}{0}
\renewcommand{\thetable}{S\arabic{table}}
\renewcommand{\theHtable}{S\arabic{table}}

\section{Description of MutAntiGen Input Parameters}
\label{S-sec:mutantigendetails}

In the MutAntiGen ABM \citep{koelle2015,bedford2012}, agents may be born, may die, may have contact with other hosts (which may or may not lead to transmission), and may recover from an infection.  Each individual carries information on all viruses by which they have already been infected, as well as information about any current viral infection.  The information on all current and past viral infections includes the antigenic type, the cross-immunity between antigenic types, the antigenic distance between the challenging strain and the host's history of encountered strains, probability of infection, and any immunities.  Viruses may mutate in antigenic phenotype, which can create turnover and outbreak in the population when there is little cross-immunity in a population to the emergence of a new phenotype.
Viruses also mutate in a ``load'' phenotype that affects their absolute fitness, independent of immunity in the host population.

We will now discuss in more detail the parameters that were set fixed in the MutAntiGen ABM, the parameters over which we sampled, and the justifications behind the ranges for the sampled parameters.  The list of possible parameters for the MutAntiGen ABM, their descriptions, and what they were set for our simulation runs, are as follows.  Parameters with an astrix were sampled according to the sampling procedures described in the main paper and according to Table \ref{tab:rates}(and therefore are not given a set value here).

\begin{enumerate}

\item \textsc{burnin}: Number of days to wait before logging output. This parameter was set to 0.

\item \textsc{endDay}: Total number of days to simulate. This parameter was set to 3650.

\item \textsc{printStep}: Interval in days at which output is written to \texttt{out.timeseries}. This parameter was set to 7.

\item \textsc{tipSamplingStartDay}: Day on which sampling for the phylogenetic tree begins, delayed to avoid basal multifurcation at time zero. This parameter was set to 56.

\item \textsc{tipSamplingEndDay}: Day on which sampling for the phylogenetic tree ends. This parameter was set to 3640.

\item \textsc{tipSamplingRate}: Number of samples stored per deme per day. This parameter was set to 1.

\item \textsc{tipSamplesPerDeme}: Maximum number of samples allowed per deme. This parameter was set to 100000.

\item \textsc{tipSamplingProportional}: Indicator of whether sampling is proportional to prevalence when multiple demes are present. This parameter was set to true.

\item \textsc{treeProportion}: Proportion of sampled tips used in tree reconstruction. This parameter was set to 1.

\item \textsc{diversitySamplingCount}: Number of samples drawn to compute diversity, Ne$\tau$, and serial interval. This parameter was set to 1000.

\item \textsc{netauWindow}: Size of the time window, in days, over which Ne$\tau$ is calculated. This parameter was set to 100.

\item \textsc{repeatSim}: Indicator of whether the simulation is repeated until \textsc{endDay} is reached. This parameter was set to false.

\item \textsc{immunityReconstruction}: Indicator of whether immunity reconstruction output is written to \texttt{out.immunity}. This parameter was set to false.

\item \textsc{memoryProfiling}: Indicator of whether memory profiling is enabled, requiring a Java agent. This parameter was set to false.

\item \textsc{yearsFromMK}: Number of years considered as present when calculating MK statistics. This parameter was set to 1.0.

\item \textsc{pcaSamples}: Indicator of whether the virus tree is rotated and flipped using PCA. This parameter was set to false.

\item \textsc{detailedOutput}: Indicator of whether detailed host and virus output files are written to enable checkpointing. This parameter was set to false.

\item \textsc{restartFromCheckpoint}: Indicator of whether the population is loaded from a previous checkpoint. This parameter was set to false.

\item \textsc{hostImmuneHistorySampleCount}: Number of host immune histories sampled when computing mean host immunity. This parameter was set to 10000.

\item \textsc{fitSampleCount}: Number of viral fitness samples collected. This parameter was set to 100.

\item \textsc{printFitSamplesStep}: Interval in days at which viral fitness samples are printed. This parameter was set to 1000.

\item \textsc{demeCount}: Number of demes in the metapopulation. This parameter was set to 1.

\item \textsc{demeNames}: Names assigned to each deme. This parameter was set to \texttt{"tropics"}.

\item $\textsc{initialNs}^*$: Initial population sizes of each deme.

\item \textsc{birthRate}: Per-individual daily birth rate. This parameter was set to 0.000091.

\item \textsc{deathRate}: Per-individual daily death rate. This parameter was set to 0.000091.

\item \textsc{swapDemography}: Indicator of whether overall population size is kept constant. This parameter was set to true.

\item \textsc{initialI}: Initial number of infected individuals. This parameter was set to 7406.

\item \textsc{initialDeme}: Index of the deme in which infection is initiated. This parameter was set to 1.

\item \textsc{initialPrR}: Proportion of the population with prior immunity to the initial strain. This parameter was set to 0.5088.

\item $\textsc{beta}^* (\beta)$: Transmission rate in contacts per individual per day. 

\item $\textsc{nu}^* (\nu)$: Recovery rate in recoveries per individual per day. 

\item \textsc{betweenDemePro}: Relative transmission rate between demes compared to within-deme transmission. This parameter was set to 0.0000.

\item \textsc{externalMigration}: Additional infections contributing to the force of infection. This parameter was set to 200.0.

\item \textsc{transcendental}: Indicator of whether a general recovered class is included. This parameter was set to false.

\item \textsc{immunityLoss}: Rate of immunity loss from recovered to susceptible per individual per day. This parameter was set to 0.0.

\item \textsc{initialPrT}: Initial fraction of the population in the general recovered class. This parameter was set to 0.0.

\item \textsc{backgroundImmunity}: Indicator of whether the population starts with background immunity to the current strain. This parameter was set to false.

\item \textsc{backgroundDistance}: Antigenic distance of background immunity from the initial strain. This parameter was set to 0.2.

\item \textsc{demeBaselines}: Baseline seasonal forcing for each deme. This parameter was set to \texttt{1.0}.

\item $\textsc{demeAmplitudes}^*$: Amplitude of seasonal forcing for each deme.

\item \textsc{demeOffsets}: Seasonal phase offsets for each deme relative to the year. This parameter was set to \texttt{0.0}.

\item \textsc{phenotypeSpace}: Representation of phenotype space used in the model. This parameter was set to \texttt{"mutLoad"}.

\item $\textsc{lambda}^* (\lambda)$: Deleterious mutation rate per genome per transmission. 

\item $\textsc{mutCost}^*$: Fitness cost associated with deleterious mutations. 

\item \textsc{probLethal}: Probability that a mutation is lethal. This parameter was set to 0.0.

\item \textsc{epsilon}: Beneficial mutation rate, scaled with population size. This parameter was set to 0.16.

\item \textsc{epsilonSlope}: Dependence of the beneficial mutation rate on mean mutational load. This parameter was set to 0.0.

\item $\textsc{epsilon\_mut}^*$: See discussion below.

\item $\textsc{initialI\_prop}^*$: See discussion below.

\item $\textsc{lambdaAntigenic}^*$: Antigenic mutation rate per transmission.

\item $\textsc{meanAntigenicSize}^*$: Mean size of antigenic mutations. 

\item \textsc{antigenicGammaShape}: Shape parameter of the gamma distribution for antigenic mutation sizes. This parameter was set to 2.0.

\item \textsc{thresholdAntigenicSize}: Minimum antigenic mutation size required for a mutation to be retained. This parameter was set to 0.012.

\item \textsc{antigenicEvoStartDay}: Day on which antigenic mutations begin to occur. This parameter was set to 0.

\item \textsc{cleanUpDistance}: Antigenic distance beyond which strains are removed from the simulation. This parameter was set to 0.2.

\item \textsc{demoNoiseScaler}: Scaling factor applied to demographic noise. This parameter was set to 0.0.

\item \textsc{muPhenotype}: Phenotypic mutation rate per individual per day. This parameter was set to 0.0.

\item \textsc{smithConversion}: Multiplier converting antigenic distance into cross-immunity. This parameter was set to 0.1.

\item \textsc{homologousImmunity}: Immunity level conferred by antigenically identical viruses. This parameter was set to 0.95.

\item \textsc{initialTraitA}: Initial value of the first dimension of host immunity. This parameter was set to -6.0.

\item \textsc{meanStep}: Mean mutation step size. This parameter was set to 0.3.

\item \textsc{sdStep}: Standard deviation of mutation step size. This parameter was set to 0.3.

\item \textsc{mut2D}: Indicator of whether mutations occur in a full two-dimensional angular space. This parameter was set to false.

\end{enumerate}

\begin{table}[H]
\centering
\small
\caption{Initial parameter ranges of the Latin Hypercube Sample}
\label{tab:rates}
\resizebox{\textwidth}{!}{
\begin{tabular}{@{}lccc@{}}
\toprule
\textbf{Parameter} & \textbf{Description} & \textbf{Initial LHS Lower Bound} & \textbf{Initial LHS Upper Bound}  \\
\midrule
N & Population Size & $1.0 \times 10^{4}$ & $1.0 \times 10^{4}$ \\
demeAmplitudes & Strength of Seasonality Forcing & $0$ & $2.0 \times 10^{-1}$\\
lambdaAntigenic & Antigenic mutation rate per transmission & $8.57 \times 10^{-5}$ & $2.57 \times 10^{-3}$ \\
meanAntigenicSize & Mean effect size of antigenic mutations & $1.2 \times 10^{-3}$  & $1.2 \times 10^{-1}$ \\
lambda & Deleterious mutation rate per genome per transmission & $9.5 \times 10^{-3}$ & $4.08 \times 10^{0}$  \\
mutCost & Fitness cost associated with deleterious mutations &$8.0 \times 10^{-4}$ & $8.0 \times 10^{-2}$  \\
beta & Transmission rate in contacts per individual per day & $1.43 \times 10^{-1}$  & $2.25 \times 10^{0}$ \\
nu & Recovery rate in recoveries per individual per day & $7.14 \times 10^{-2}$  & $2.5 \times 10^{-1}$ \\
epsilon\_mut & Multiplier to the baseline scaling of the fraction of beneficial mutations & $5.0 \times 10^{-1}$  & $1.5 \times 10^{0}$  \\
initialI\_prop & Initial proportion of population size infected & $1.0 \times 10^{-4}$&  $1.0 \times 10^{-3}$ \\ 
\bottomrule
\end{tabular}
}
\end{table}

The full set of bounds for the parameters that were sampled via a Latin hs

\section{Computational Modifications to the MutAntiGen Software} \label{S-sec:mutantigen_coding}

There were several challenges to running a massive amount of MutAntiGen simulations over these parameter ranges.  First, it is possible for a given simulation to ``break" due to the current state of the system becoming untenable for the given computational resources\footnote{All computations were performed on an HPC cluster running Red Hat Enterprise Linux 8.10 with Slurm, using dual-socket Intel Xeon E5-2695 v3 x86\_64 nodes (56 hardware threads, $\sim256$ GB RAM per node) compiled with GCC 8.5.0.}.  Second, it was nearly impossible to predict how long a given simulation might take: for two simulation runs that were very similar in terms of parameter settings, one might complete fully in 30 minutes, while the other might stall out after running for 7 days.  Third, not all runs showed disease turnover and behaviors of multiple antigenic mutations.  Lastly, the original coding of MutAntiGen was not well-designed for running thousands of simulations in parallel on an HPC environment. 

\begin{figure}[H]
  \centering
  \includegraphics[width=1\textwidth]{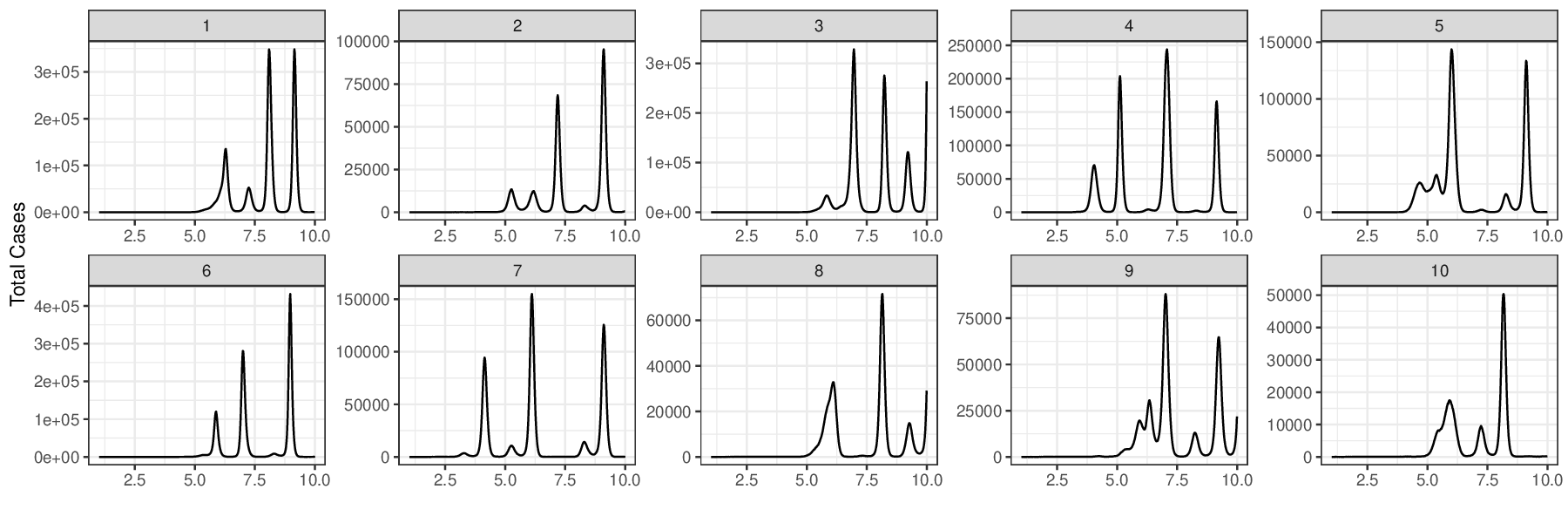}
  \caption{10 MutAntiGen output simulated from identical input parameters..}
  \label{fig:synthetic_data_replicates}
\end{figure}

Several modifications were coded and strategies taken to extract robust and diverse synthetic data via the MutAntiGen ABM and to handle the above challenges.  First, we re-coded the model to allow for multiple simulations to be run in parallel on the same HPC system; this modified version is available on this paper's GitHub page (\url{https://github.com/lanl/precog}).  This re-coding also disabled some functionality of MutAntiGen that was not necessary to these experiments and was stalling out some of the simulations.  Second, we created a ``max wall time" that would cut simulations that did not complete within a certain time.  While this cut off may bias the sampling, we did not have the computational resources to allow a given run to go for longer than 10 hours.  Lastly, we performed a modified space-filling design for the selection of parameters of the ranges of interest that incorporated subject matter expertise:
\begin{enumerate}
    \item We performed simulations over a sparse Latin-Hypercube sample (LHS) \cite{mckay1979} of size 4000 over the parameter ranges given above.  Of these 4000 samples, 815 completed.
    \item Following this initial LHS, we determined which sets of input parameters led to simulations that represented antigenic mutation and turnover in dominant antigenic type. There were 151 sets of such input parameters, which corresponded to around 18\% of the completed simulations.
    \item We performed 20 simulations at each of these input parameters, using a different seed for each input; since MutAntiGen is heavily stochastic, these replications still gave very different results.  As an example, Figure \ref{fig:synthetic_data_replicates} shows 10 of the 20 replications for a single input parameter set.
    \item Of the simulations run in replicate, 1832 completed successfully.  Around 65\% of these completed simulations exhibited antigenic mutation and turnover in dominant antigenic type. This final set was used to train the forecasting model.
\end{enumerate}

\section{Observation Model Details}
\label{S-sec:obsmodel}

Let $y_{1:T}$ be a time series of length $T$.
In this section, we describe the observation model.
That is, we describe how we stochastically convert $y_{1:T}$ into $z_{1:T'}$.
For every MutAntiGen time series $y_{1:T}$, the observation model creates 20 new time series.
The observation model has three steps: (1) scaling, (2) noise and (3) outliers.
All 20 observation model time series undergo scaling, 10 of the observation model time series undergo noise, and each of the 20 realizations has a 25\% chance to undergo outliers.
Below we detail the observation model.

\paragraph{Scaling.}
We sample a new number of time steps $T'$ for the time series $x_{1:T'}$, where
\begin{align*}
    T' &\sim \text{Uniform}(52,T)
\end{align*}
\noindent (uniform over the integers). 
We then linearly interpolate the original time series $y_{1:T}$ to be of length $T'$, resulting in a new time series $x_{1:T'}$. In \texttt{R} code, this is done via the \texttt{approx()} function:
\begin{center}
\texttt{x[1:T'] = approx(x = 1:T, y = y, xout = seq(1, T, length.out = T'))\$y}
\end{center}
\noindent This new time series $x_{1:T'}$ has the same dynamics as in $y_{1:T}$, but those dynamics can occur on a faster time scale than $y_{1:T}$.
The smaller $T'$, the faster the time scale.

\paragraph{Noise.}
Next, we add noise to $x_{1:T'}$, resulting in a new time series $v_{1:T'}$.
The process for adding noise is as follows:
\begin{align*}
  \kappa &\sim \text{Uniform}(1.5, 3.5) \\
  \epsilon_t|\kappa &\sim \text{Uniform}(\kappa^{-1}, \kappa)\\
  v_t &= x_t * \epsilon_t
\end{align*}
\noindent Note that $\kappa$ is common to the time series $x_{1:T'}$ while $\epsilon_t$ is different for each element of the time series.
The noise is multiplicative where $\epsilon_t \in [3.5^{-1}, 3.5]$.
When $\kappa$ is close to 3.5, more noise is added to the time series.
When $\kappa$ is close to 1.5, less noise is added.
If $x_{1:T'}$ is a time series that we do not add noise to, then we simply set $v_{1:T'}$ to $x_{1:T'}$.

\paragraph{Outliers.}
Time series of public health data can have high outliers due to newly discovered data dumped onto a single date or low outliers due to non-reporting. 
It is important to have training data that accurately reflects real-world data, otherwise a machine learning model --- having never seen outliers --- may assume a large leap or drop in the time series reflects a meaningful change in the time series dynamics rather than a spurious change in the data collection process.

We randomly select between 5 and 10 time steps $\in \{1,2,\ldots,T'\}$ to be outliers. 
Half of those time steps represent low outliers and the other half represent high outliers. 
We define multipliers $\lambda_t$ for all times $t \in \{1,2,\ldots,T'\}$ as follows:
\begin{align*}
  \lambda_t &= 
  \begin{cases}
      \text{Uniform}(2, 10) & \text{if } t \text{ is a high outlier}\\
      \text{Uniform}(0, 0.05) & \text{if } t \text{ is a low outlier}\\
      1 & \text{if } t \text{ is not an outlier}
  \end{cases}
\end{align*}
\noindent Finally, we define the output time series of our observation model, $z_{1:T'}$ as follows:
\begin{align*}
    z_t &= v_t*\lambda_t.
\end{align*}
\noindent 10 realizations of $z_{1:T'}$ are plotted in Figure \ref{fig:ex_synthetic_data_imperfect}.

\paragraph{Application.}
To create the synthetic total cases training data, we apply the observation model to the approximately 1,800 MutAntiGen outputs where each input is recycled 20 times making 20 new $z_{1:T'}$ time series. 
To create the synthetic variant attributable cases training data, we sample up to ten of the variant time series and, for each one of them, we apply the scaling plus outlier routines (making one new $z_{1:T'}$ time series) and separately apply the scaling plus noise plus outlier routines (making another new $z_{1:T'}$).
That is, each variant time series makes two new $z_{1:t'}$ time series.
The total number of time series for training, synthetic total cases or synthetic variant attributable cases, results in about 36,000 time series (refer to Table \ref{tab:train_ts} for specifics).

\section{Transformer Model Details}
\label{S-sec:model_details}

\paragraph{Data and objective.}
We consider a collection of univariate time series, $y_{1:T}$, each provided as observations of nonnegative counts $y_t$ (``cases'') indexed by time $t$. Our goal is probabilistic forecasting: given the most recent $C=20$ observations, we predict the distribution of the next $H=4$ observations. Forecast uncertainty is represented via predictive quantiles at levels 
$$\tau \in \mathcal T = \{0.0005, 0.005, 0.01, 0.025, 0.05, 0.1, \ldots, 0.9, 0.95, 0.975, 0.99, 0.995, 0.9995\}$$
\noindent where $\tau \in [0,1]$ and $|\mathcal T| = 27$.

\paragraph{Training examples (rolling windows).}
Training uses randomly sampled rolling windows. For a series $y_{1:T}$, we repeatedly sample a start time $t$ such that a contiguous block of length $C+H$ exists, and define the input and outputs (targets) as
\begin{equation*}
  \mathbf{y}^{\text{in}}_t = \{ y_{t-C+1}, y_{t-C+2},\ldots,y_{t} \},
  \qquad
  \mathbf{y}^{\text{out}}_t = \{y_{t+1},\ldots,y_{t+H}\}.
\end{equation*}
Mini-batches are formed by sampling a series with probability proportional to its number of available windows, then sampling a window uniformly within that series.

\paragraph{Per-window normalization and rescaling.}
To stabilize training across series with different magnitudes of case counts, each input window is normalized by its own maximum
\begin{align*}
  m_t = \max(\mathbf{y}^{\text{in}}_t), \qquad
  \mathbf{z}^{\text{in}}_t = \mathbf{y}^{\text{in}}_t / m_t,
  \qquad
  \mathbf{z}^{\text{out}}_t = \mathbf{y}^{\text{out}}_t / m_t,
\end{align*}
with the convention that if $m_t = 0$ (or is non-finite) we do not rescale that window. 
The model operates on $\mathbf{z}^{\text{in}}_t$ and outputs forecasts on the normalized scale. 
Predictions are returned to the original scale by multiplying by $m_t$.

\paragraph{Input perturbation (augmentation).}
To improve robustness to observation noise and related data-generating irregularities, each sampled training example is duplicated with a perturbed input context:
\begin{align*}
  \tilde{y}_{t} = y_t * u_t,
  \qquad
  u_t \overset{\text{iid}}{\sim} \mathrm{Uniform}(0.85, 1.15).
\end{align*}
Thus, $\tilde{\mathbf{y}}^{\text{in}}_t = \{ \tilde{y}_{t-C+1}, \tilde{y}_{t-C+2},\ldots,\tilde{y}_{t} \}$ is a perturbed version of $\mathbf{y}^{\text{in}}_t$.
The future targets $\mathbf{y}^{\text{out}}_t$ are unchanged. 
What this seeks to accomplish is a more stable learning representation.
By providing the model with two distinct inputs that map to the same output, the model should be better positioned to learn stable underlying representations and less likely to learn spurious ones.
This yields an effective batch size that is twice the nominal batch size.

\paragraph{Model architecture.}
We represent the forecasting function with a compact transformer encoder. 
Each scalar input in $\mathbf{z}^{\text{in}}_t$ is mapped to a $d$-dimensional embedding ($d=256$) via a learned linear projection, and a learned positional embedding is added for the $C$ time positions. 
The embedded sequence is processed by a transformer encoder with $L=2$ layers and $n_{\text{head}}=4$ attention heads (feedforward dimension 512). The final hidden state (at the most recent input time point) is mapped through a linear readout to produce $H \times |\mathcal T|$ outputs.

\paragraph{Quantile parameterization.}
The model outputs predictive quantiles $\hat{z}_{h,\tau}$ for each horizon $h=1,\ldots,H$ and quantile level $\tau \in \mathcal T$. To prevent quantile crossing, quantiles are constructed using a base quantile plus nonnegative increments:
\begin{align*}
  \hat{z}_{h,\tau_1} = a_{h,1}, \qquad
  \hat{z}_{h,\tau_k} = \hat{z}_{h,\tau_1} + \sum_{j=2}^{k} \mathrm{softplus}(a_{h,j}),
  \quad k=2,\ldots,|\mathcal T|,
\end{align*}
where $a_{h,j}$ are unconstrained network outputs and $\mathrm{softplus}(x)=\log(1+e^x)$ ensures positivity of the increments. Forecasts on the original scale are obtained as $\hat{y}_{h,\tau} = m_t \, \hat{z}_{h,\tau}$.

\paragraph{Loss function (pinball loss).}
Parameters are estimated by minimizing the mean quantile (pinball) loss over horizons and quantile levels. For observation $y_{t+h}$ and predicted quantile $\hat{y}_{t+h,\tau}$,
\begin{align*}
  \rho_{\tau}(y_{t+h} - \hat{y}_{t+h,\tau})
  = \max\!\bigl(\tau (y_{t+h} - \hat{y}_{t+h,\tau}),\, (\tau-1)(y_{t+h} - \hat{y}_{t+h,\tau})\bigr),
\end{align*}
and the training objective is the average of $\rho_{\tau}$ across all $(h,\tau)$ and all examples in each mini-batch.

\paragraph{Optimization, validation, and model selection.}
We optimize the model using Adam with a cosine-decayed learning rate schedule (from $5\times 10^{-4}$ to $5\times 10^{-5}$) over the prescribed number of weight updates (we used 97,656). 
Validation uses a fixed set of randomly sampled windows (held constant by a fixed seed).
We maintain an exponential moving average (EMA) of parameters with smoothing coefficient $\alpha_{\mathrm{EMA}}=0.98$ and select the final model as the EMA checkpoint with the lowest validation loss.

\section{Bootstrapping Details}
\label{S-sec:bootstrapping}
\begin{figure}[H]
  \centering
  \includegraphics[width=1\textwidth]{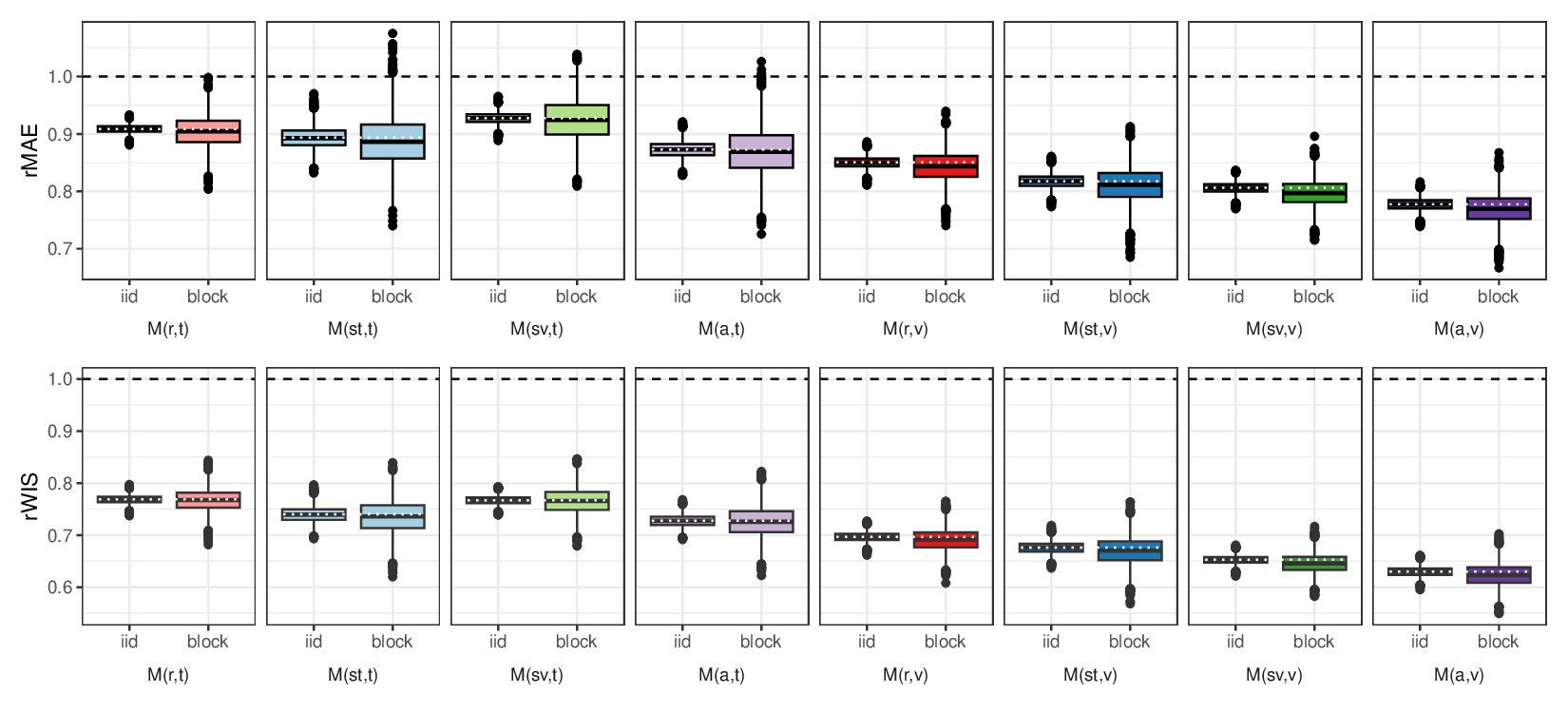}
  \caption{Boxplots for 5,000 bootstrap samples for rMAE and rWIS from the blocked bootstrap procedure (block) taking into account state and forecast date groupings and the vanilla bootstrap procedure (iid) that does not address groupings. Wider uncertainties are observed when taking account of the grouping variables. The white, dotted horizontal line is the point estimate for rMAE and rWIS.}
  \label{S-fig:bscomp}
\end{figure}

In this study, nine different models made 27,540 forecasts (51 states/territories $\times$ 135 forecast dates $\times$ 4 forecasts horizons).
Our goal is to determine which models produce forecasts that are statistically significantly better than other models. 
As a zeroth order pass, we can compute the metric of interest (e.g., MAE or WIS) for each model over the entire set of forecasts, and compare the numbers. 
Whichever model produced the best number is the best forecasting model.
This, however, does not take into account uncertainty. 
To get an estimate of uncertainty, we turn to bootstrapping \cite{efron1994introduction}.
This requires us to resample the 27,540 forecasts with replacement.
To deal with the groupings (e.g., states and forecast dates), we first sample states with replacement and then, for a sampled state, sample nine different blocks of time (where a block of time is 15 consecutive weeks of forecast dates).
For each state/time block, we keep all horizons and all models.
This results in a resampling of 27,540 forecasts per model (51 resampled states $\times$ 9 resampled time blocks $\times$ 15 dates per block $\times$ 4 horizons $=$ 27,540 forecasts).
For each resampled data set, we compute each model's MAE and WIS.
We repeat this data set resampling procedure 5,000 times.
The results of the block bootstrap procedure (block) and the vanilla bootstrap procedure which ignores groupings (iid) are shown in Figure \ref{S-fig:bscomp}.
Taking into account the grouping variables in the block bootstrap procedure results in wider uncertainty intervals. 
The 95\% confidence intervals presented in the main text were derived following this block bootstrap procedure.

\end{document}